\documentclass[showpacs,preprint]{revtex4}
\usepackage{graphicx}
\usepackage{dcolumn}
\usepackage{amsmath}
\usepackage[latin1]{inputenc}
\usepackage{graphicx, psfrag}
\usepackage{amssymb}
\usepackage[colorlinks=true, citecolor=blue, urlcolor = blue, linkcolor= red, bookmarks=true]{hyperref}
\usepackage{float}
\usepackage{amsmath}
\usepackage{amsfonts}
\usepackage{dcolumn}
\usepackage{hyperref}
\usepackage{subfigure}
\usepackage{pgfplots}
\usepackage{epstopdf}
\def \beq{\begin{equation}}
\def \eeq{\end{equation}}
\def \bse{\begin{subequations}}
\def \ese{\end{subequations}}
\def \bea{\begin{eqnarray}}
\def \eea{\end{eqnarray}}
\def \bem{\begin{displaymath}}
\def \eem{\end{displaymath}}
\def \bem{\begin{pmatrix}}
\def \eem{\end{pmatrix}}
\def \beb{\begin{bmatrix}}
\def \eeb{\end{bmatrix}}
\def \bc{\begin{center}}
\def \ec{\end{center}}

\makeatletter
\def\btt#1{\texttt{\@backslashchar#1}}
\DeclareRobustCommand\bblash{\btt{\@backslashchar}} \makeatother

\makeatletter
\def\btt#1{\texttt{\@backslashchar#1}}
\DeclareRobustCommand\bblash{\btt{\@backslashchar}} \makeatother

\begin{document}
	
\title[]{Thermodynamics of ${P}$-${V}$ criticality in $d$-dimensional AdS black holes surrounded by a perfect fluid in Rastall theory}
 \author{Md Sabir Ali $^{a,b}$} \email{alimd.sabir3@gmail.com}
 \affiliation{$^{a}$ Department of Physics, Mahishadal Raj College, West Bengal 721628, India}
\affiliation{$^{b}$ Center for
		Theoretical Physics, Jamia Millia Islamia, New Delhi 110025, India}
\begin{abstract} 
We study a $d$-dimensional Anti-de Sitter (AdS) black hole surrounded by a static anisotropic quintessence field within the framework of Rastall theory. The solution is characterized by several parameters including its mass ($M$), field structure parameter ($N_s$), Rastall coupling parameter ($\psi$), and the cosmological constant ($\Lambda$). Our objective here is to identify the $d$-dimensional black holes within the framework of Rastall theory, exploring special cases, e.g., a cosmological constant,  dust, radiation, and quintessence fields. In addition to this, we derive the effective equation of state parameter, denoted as $\omega_{eff}$. Next, we investigate the thermodynamics for $P$-$V$ criticality and phase transitions in the extended phase space of black hole thermodynamics. For a specific set of values for the Rastall coupling parameter, we numerically plot isothermal and isobaric curves in the reduced parameter space. We also compute the specific heat at constant pressure ($C_P$ ), volume expansion coefficient ($\alpha$), and isothermal compressibility ($\kappa_T$) to deepen our understanding of the analogy between the thermodynamics of Rastall AdS black holes and that of a liquid-gas system. Our investigations indicate that the dimension of spacetime and the Rastall coupling parameter significantly influence the critical nature of phase transitions. By utilizing the expressions for $C_P$,  $\alpha$, and $\kappa_T$, we derive the Ehrenfest equations and perform an analytical investigation of phase transitions at their critical points. These results allow us to compare the thermodynamics of AdS black holes with liquid-gas systems, which closely mimics the behavior of van der Waals (vdWs) gases. 

\end{abstract}


\maketitle
\section{Introduction}
Einstein's general relativity (GR) successfully describes many predictions within its regime of validity and has become the most widely accepted and successful theory of gravity. In the framework of GR, geometry of the spacetime and matter fields are minimally coupled, resulting in the covariant conservation relation of the energy-momentum tensor (EMT). However, when there is non-minimal coupling between geometry and matter fields, their mutual interactions can significantly influence each other \cite{Nojiri:2004bi, Allemandi:2005qs, Koivisto:2005yk, Bertolami:2007gv, Harko:2014gwa}. This non-minimal coupling leads to a violation of covariant conservation of the EMT of matter fields \cite{Rastall:1973nw, Rastall:1976uh}. Traditionally, the idea of covariant conservation based on spacetime symmetries has been applied primarily in Minkowaski flat spacetime and/or in the weak-field regime of gravity. However, in the strong gravitational field regime, debates continue regarding the true nature of spacetime geometry and the covariant conservation relation. At this point, Rastall \cite{Rastall:1973nw, Rastall:1976uh} proposed a phenomenological model in which covariant conservation of EMT is violated. This model is expressed in the form $T{^{\mu\nu}{_{;\mu}}}=\lambda\;R{^{;\nu}}$, where $R$ is the Ricci scalar and $\lambda$ is the Rastall coupling parameter. Here, the parameter $\lambda$ quantifies the potential deviations of Rastall theory from GR and indicates a curvature-matter coupling in a non-minimal manner. Interestingly, all electrovacuum solutions to Einstein's GR are also solutions of Rastall theory, and asymptotically, both theories approach Mankowski spacetime. However, the non-vacuum solutions in Rastall gravity incorporate the Rastall parameter, making them significantly different from their corresponding solutions in GR. This difference contributes to the aesthetic richness of Rastall gravity\cite{Heydarzade:2017wxu}. \\
In recent years, the Rastall theory has attracted great attention, and a rich diversified research dedicated to it is available in the literature. Some of these include phenomenological results related to both astrophysical \cite{Bedford:1988zv, Heydarzade:2017wxu, Heydarzade:2016zof, Bronnikov:2016odv, Bronnikov:2017pmz} and cosmological consequences \cite{AbdelRahman:1997sk, AlRawaf:1994pn, De:1999vh, Arbab:2002tg, Batista:2011nu, Batista:2010nq, Batista:2012hv}. Notable examples include the static, cylindrically symmetric black hole solutions to Rastall gravity coupled to the $U(1)$ Abelian-Higgs model, which takes into account quantum effects in a curved spacetime characterized by a phenomenological manner \cite{deMello:2014hra}. Recently, a variety of black hole solutions based on Rastall theory have been explored. Among these are spherically symmetric black hole solutions \cite{Heydarzade:2017wxu, Heydarzade:2016zof}, rotating black holes \cite{Kumar:2017qws, Xu:2017vse}, and studies of their shadow properties \cite{Kumar:2017vuh}. Additionally, many solutions have contributed to the understanding of the thermodynamics and other theoretical aspects of black holes \cite{Licata:2017rfx, Carames:2014twa, Salako:2016ihq}. Some research has also focused on comparing the Rastall theory with standard GR \cite{Visser:2017gpz, Moradpour:2017tbp}. \\
The essence of a dark energy fluid known as quintessence has been incorporated into the black hole solution proposed by Kiselev \cite{Kiselev:2002dx}. Black hole solutions surrounded by the quintessence field have been generalized to higher dimensions in GR \cite{Chen:2008ra} as well as in other theories of gravity \cite{Ghosh:2017cuq}. Additionally, quintessence black hole solutions have been extended to Rastall theory \cite{Heydarzade:2017wxu}. Recently, a solution in $d$-dimensions to Rastall theory with the presence of perfect fluid has been obtained \cite{MoraisGraca:2017hrf} and further extended to charged anti-de Sitter (AdS) spaces \cite{Lin:2018coh}. \\

The purpose of this paper is to conduct an analytical study of the extended phase space thermodynamics of a $d$-dimensional AdS black holes within the context of Rastall theory, focusing on their $P$-$V$ criticality. The concept of AdS black holes thermodynamics originated from the pioneering work of Hawking and Page \cite{Hawking:1982dh}, which described a first-order phase transition between the Schwarzschild AdS black holes and the thermal AdS spaces. Since then, interest in the phase transitions of AdS black holes has continued to grow. Black holes in AdS spacetimes exhibit characteristics of real thermodynamic systems, as they incorporate pressure, volume, and temperature-- factors essential for studying the thermodynamic phenomena and their rich phase structures. These structures incorporate van der Waals (vdWs) phase transitions, $P$-$V$ criticality, reentrant phase transitions, triple points, isolated critical points, and superfluidity \cite{Kubiznak:2012wp, Gunasekaran:2012dq, Altamirano:2013ane, Kastor:2009wy, Altamirano:2014tva, Altamirano:2013uqa, Frassino:2014pha, Wei:2014hba, Dolan:2014vba, Wei:2015ana, Hennigar:2016xwd}. These considerations motivate us to study the AdS black holes as extended thermodynamic systems. Additionally, $d$-dimensional charged AdS black holes also exhibit vdWs phase transitions \cite{Kastor:2009wy, Altamirano:2014tva, Altamirano:2013uqa, Frassino:2014pha, Wei:2014hba, Dolan:2014vba, Wei:2015ana, Hennigar:2016xwd}. In a reduced parameter space, it has been found that the critical phenomena are independent of charge parameter \cite{Kastor:2009wy, Altamirano:2014tva, Altamirano:2013uqa, Frassino:2014pha, Wei:2014hba, Dolan:2014vba, Wei:2015ana, Hennigar:2016xwd}. The critical phenomena of the AdS black holes have also been extended in some modified gravity theories, such as massive gravity theory \cite{Hendi:2015hoa}, power-law Maxwell field \cite{Hendi:2016usw}, Born-Infeld theory \cite{Hendi:2014kha}, and regular AdS black holes \cite{Ali:2019myr}. Therefore, it is interesting to study the thermodynamic phase transitions and $P$-$V$ criticality of $d$-dimensional AdS black holes in the presence of a quintessence field within Rastall gravity. The impact of the Rastall parameter on thermodynamic quantities is rather important, as it has noteworthy contributions to these properties.\\
The off-shell Gibbs free energy approaches have been utilized to establish important connections to thermodynamic properties of black holes in asymptotically AdS spacetime. In particular, such explorations have been successfully employed for Schwarzschild-AdS black holes, thereby drawing on their inherent relationship with Hawking-Page phase transitions. A similar procedure are then applied to examine the thermodynamic properties of AdS black in massive gravity scenarios \cite{Li:2020khm}. We express the off-shell Gibbs free energy as a function of the horizon radius, which serves as an order parameter. Although significant progress has been made in shaping the off-shell Gibbs free energy, a complete kinematic description is still in development. Since its first application to Schwarzschild-AdS black hole systems, the off-shell Gibbs free energy has been explored for a wide variety of AdS black hole spacetime in both Einstein's general relativity and alternative theories of gravity. Recently, there has been a surge of interest in analyzing the off-shell Gibbs free energy approaches. These procedures have been utilized in Reissner-Nordstr$\ddot{o}$m-AdS black hole systems, charge-neutral/charged AdS black holes in Einstein-Gauss-Bonnet theory \cite{Li:2020nsy, Wei:2020rcd, Li:2020spm}. A similar analysis has been extended to analyze the dynamics of triple point of a higher dimensional charged Gauss-Bonnet-AdS system \cite{Wei:2021bwy} and subsequently extended to charged AdS systems in a dark energy environment \cite{Lan:2021crt}. The method has been utilized for AdS black hole solutions when general relativity is minimally coupled to nonlinear electromagnetic theory including the Euler-Heisenberg-AdS black hole spacetime \cite{Dai:2022mko, Kumara:2021hlt, Ali:2023uwh}. Furthermore, similar analyses have been conducted for the rotating spacetime, as can be seen for Kerr-AdS black holes \cite{Yang:2021ljn}. \\

Our paper is organized as follows. In the next section, we provide a brief discussion of $d$-dimensional AdS black holes in the presence of a static, anisotropic quintessence field within the context of Rastall theory. We also explore specific cases including dust, radiation, and quintessence matter. In Sec.~\ref{decomp}, we demonstrate that an anisotropic quintessence field can be decomposed into a perfect fluid component and electromagnetic field components, provided that the null energy condition is satisfied. In Sec.~\ref{horizon_struc}, we provide a simple and concise discussion on the horizon structure of black holes. In Sec.~\ref{thermodynamics}, we evaluate the thermodynamic quantities associated with the black holes, calculating expressions for thermodynamic pressure and volume as relevant variables. In Sec.~\ref{PV_criticality}, we employ the classical Ehrenfest equations to analytically examine the AdS black holes at the critical points of $P$-$V$ criticality. A little has been put to investigate a qualitative analysis on the off-shell Gibbs free energy approaches in Sec.~\ref{free_energy_LS}. Finally, we conclude the paper in Sec.~\ref{conclusion}. 

\section{$d$-dimensional AdS black holes spacetime surrounded by an anisotropic quintessence field}
The non-conservation of EMT in the strong gravity regime, as proposed by Rastall \cite{Rastall:1973nw, Rastall:1976uh}, leads to the relation $T^{\mu\nu}{_{;\mu}}=\lambda\; R{^{;\nu}}$, where $R$ is the Ricci scalar and $\lambda$ is the Rastall parameter amounting for the potential deviation from standard GR. This assumption results in modifications to Einstein's field equations, which can be written as follows
\begin{eqnarray}\label{EFEs}
G_{\mu\nu}=\kappa \left(T_{\mu\nu}-\lambda g_{\mu\nu}R\right)
\end{eqnarray}
where $\kappa$ is a coupling constant related to the Newton's gravitational constant, $G_N$, via $\kappa=8\pi G_N$. 
If we include the negative cosmological constant $\Lambda$, the field equations (\ref{EFEs}) can be rewritten as
\begin{eqnarray}\label{EFE1s}
G_{\mu\nu}+\Lambda g_{\mu\nu}=\kappa \left(T_{\mu\nu}-\lambda g_{\mu\nu}R\right).
\end{eqnarray}
The trace of the Eq.~(\ref{EFE1s}) leads to have the following form
\begin{eqnarray}
\label{EFE2s}
G_{\mu\nu}+\Lambda g_{\mu\nu}=\kappa \tilde{T}_{\mu\nu},
\end{eqnarray}
where $\tilde{T}_{\mu\nu}$ is the effective EMT having the form
\begin{eqnarray}
\tilde{T}_{\mu\nu}=T_{\mu\nu}-\frac{\psi\left(T-\tilde{\Lambda}d \right)}{\left(\psi-1/2 \right)d+1}g_{\mu\nu}
\end{eqnarray}
in which $d$ is the spacetime dimension, $\tilde{\Lambda}=\Lambda/\kappa$ and $T_{\mu\nu}$ is the EMT for surrounding quintessence field with $T$ being its trace. Here and henceforth, we consider $\kappa\lambda=\psi$. The EMT $T_{\mu\nu}$ in $d$-dimensional spacetime for the surrounding quintessence field is expressed as follows \cite{Chen:2008ra, Ghosh:2017cuq}, 
\begin{eqnarray}\label{EMT1}
{T}^{t}{_t}&=&{T}^{r}{_r}=-\rho_s,
\end{eqnarray}
and for the angular components, we have
\begin{eqnarray}\label{EMT2}
T^{\theta_{1}}{_{\theta_{1}}}&=&T^{\theta_{2}}{_{\theta_{2}}}=...=T^{\theta_{d-2}}{_{\theta_{d-2}}}\nonumber\\
&=&\frac{\rho_s}{d-2}\left[\left(d-1\right)\omega_s+1\right],
\end{eqnarray}
where $\omega_s$ and $\rho_s$ are, respectively, the equation of state parameter and energy density of the surrounding quintessence field. 
With these expressions of $T^{\mu}{_\nu}$, the components of effective EMT $\tilde{T}^{\mu}{_\nu}$ read as
\begin{eqnarray}
\label{effectiveEFEs}
\tilde{T}^{t}{_t}&=&\tilde{T}^{r}{_r}=-\frac{\Bigg[\left(2\psi-1\right)\left(d-1\right)^2\left(1+\omega_s\right)-\left(d-2\right)^2\Bigg]\rho_s-2\tilde{\Lambda}d(d-2)}{(d-2)((2\psi-1)d+2)},\nonumber\\
\tilde{T}{^{\theta{_1}}{_{\theta{_1}}}}&=&\tilde{T}{^{\theta{_2}}{_{\theta{_2}}}}=...=\tilde{T}{^{\theta{_{d-2}}}{_{\theta{_{d-2}}}}},\nonumber\\
&=&\frac{\Bigg[\left(2\psi-1\right)\left(d-1\right)\left(1+\omega_s\right)+1\Bigg]\rho_s-2\tilde{\Lambda}d(d-2)}{(d-2)((2\psi-1)d+2)}.\nonumber\\
\end{eqnarray}
The solutions for black holes in a $d$-dimensional static spherically symmetric spacetime surrounded by a quintessence field have been analyzed \cite{MoraisGraca:2017hrf} and subsequently extended to charged AdS spacetimes \cite{Lin:2018coh}. In this discussion, we focus on the $d$-dimensional static spherically symmetric black holes surrounded by a quintessence field in AdS spacetime \cite{MoraisGraca:2017hrf, Lin:2018coh}. The line element for a $d$-dimensional spherically symmetric Schwarzschild-Tangherlini-like solution reads
\begin{eqnarray}\label{metric}
ds^2&=&-f(r)dt^2+\frac{dr^2}{f(r)}+r^2d\Omega_{d-2}^2,
\end{eqnarray}
where d$\Omega_{d-2}^2$ is the line element of a unit sphere of dimension $(d-2)$ expressed as
\begin{eqnarray}
d \Omega^{2}_{d-2} = d \theta^{2}_{1} + \sum^{d-2}_{i=2}  \left[ \prod^{i}_{j=2} \sin^2 \theta_{j-1} \right] d \theta^{2}_{i}.
\end{eqnarray}
The field equations (\ref{EFE2s}) together with the metric (\ref{metric}) lead to the following independent equations.
\begin{eqnarray}\label{eos1}
\frac{d-2}{2r^2}\left[rf{^\prime}+\left(d-3\right)\left(f-1\right)\right]+\Lambda &=&\kappa\tilde{T}^{t}{_t}
\end{eqnarray}
\begin{eqnarray}\label{eos2}
\frac{1}{2r^2}\left[r^2f^{^\prime{^\prime}}+2\left(d-3\right)rf{^\prime}+\left(d-3\right)\left(d-4\right)\left(f-1\right)\right]+\Lambda &=&\kappa\tilde{T}{^{\theta{_{d-2}}}{_{\theta{_{d-2}}}}}
\end{eqnarray}
From Eqs.~(\ref{effectiveEFEs})-(\ref{eos2}) we get a master differential equation to solve $f$:
\begin{eqnarray}\label{master}
&&\left[1-2\psi\frac{\left(d-1\right)\left(1+\omega_s\right)}{d-2}\right]r^2f{^\prime{^\prime}}+\left[\left(d-1\right)\omega_s+\left(2d-5\right)-4\psi{\left(d-1\right)\left(1+\omega_s\right)}\right]rf{^\prime}\nonumber\\
&+&\left(d-2\right)\left[\left(d-1\right)\omega_s+d-3-2\psi{\left(d-1\right)\left(1+\omega_s\right)}\right](f-1)\nonumber\\
&=&\frac{2(d-1)(1+\omega_s)}{(d-2)}{\Lambda r^2}\nonumber\\
\end{eqnarray}
Eq.~(\ref{master}) has a solution of the form
\begin{eqnarray}\label{metricfunads}
f(r)&=&1-\frac{m}{r^{d-3}}-\frac{N_s}{r^{\xi}}+\frac{r^2}{l^2},
\end{eqnarray}
where $m$ and $N_s$ are integration constants that are related to the black hole mass and the density of the surrounding quintessence field, respectively. The parameter $m$ is related to the Arnowitt-Deser-Misner (ADM) mass $M$ of the black hole as 
\begin{eqnarray}
M=\frac{(d-2)}{16\pi }\Omega_{d-2}m,\,\,\,\,\,\Omega_{d-2}=\frac{2\pi^{\frac{d-1}{2}}}{\Gamma(\frac{d-1}{2})}.
\end{eqnarray}\label{Mass}
Additionally, we use the parameter $\xi$ in shorthand notation, which is explicitly written as
\begin{eqnarray}
\xi=\frac{\left(d-3\right)+\left(d-1\right)\omega_s-2\psi\left(d-1\right)\left(1+\omega_s\right)}{1-2\psi\left(\frac{d-1}{d-2}\right)\left(1+\omega_s\right)}.
\end{eqnarray}
In Rastall theory, the definition of the curvature radius $l$ is modified. The rationale for designating the term $l$ as the curvature radius in Eq.~(\ref{metricfunads}) is explained in the follwoing discussion. Although this definition resembles that in Einstein's gravity theory, the analytical expression differs significantly because the definition of the negative cosmological constant incorporates the Rastall coupling parameter, $\psi$. The $\mathrm {AdS} _{d}$ metric with curvature radius $l$ represents one of the maximal symmetric $d$-dimensional spacetimes. Its geometric properties can be described as follows: the Riemann tensor for such spacetime is represented as
\begin{eqnarray}
\label{riemann}
R_{\mu\nu\alpha\beta}=-\frac{1}{l^2}\left(g_{\mu\alpha}g_{\nu\beta}-g_{\mu\beta}g_{\nu\alpha}\right).
\end{eqnarray}
This alllows us to derive the Ricci tensor as
\begin{eqnarray}
\label{ricci}
R_{\mu\nu}=-\frac{(d-1)}{l^2}g_{\mu\nu},
\end{eqnarray}
which consequently leads to the Ricci scalar given by 
\begin{eqnarray}
\label{ricciscalar}
R=-\frac{d(d-1)}{l^2}
\end{eqnarray}
Using Einstein's field tensor $G_{\mu\nu}$ along with Eqs.~(\ref{ricci}) and (\ref{ricciscalar}), we obtain
\begin{eqnarray}
\label{G_munu}
G_{\mu\nu}=\frac{(d-1)(d-2)}{2l^2}.
\end{eqnarray}
The field tensor $G_{\mu\nu}$, Eq.~(\ref{G_munu}), and the Eqs.~(\ref{EFE1s}) with $T_{\mu\nu}=0$ leads us to have
\begin{eqnarray}
\Lambda=-\frac{(d-1)((1-2\psi)d-2)}{2l^2}.
\end{eqnarray}
We have demonstrated from the basic definition that expression for the cosmological constant is different in Rastall theory. This change is solely attributed to the Rastall coupling parameter $\psi$. When we set $\psi=0$, we obtain the expression for the cosmological constant as defined in general relativity and other modified theories of gravity. The significance of the Rastall coupling parameter deserves further investigation. It is important to mention that the effects of the cosmological constant are diminished by the presence of the Rastall coupling parameter, while the curvature radius is increased. \\
The energy density $\rho_s$ can be obtained by substituting $f(r)$ from Eq.~(\ref{metricfunads}) into Eq.~(\ref{eos2}), which leads to the following expression
\begin{eqnarray}
\rho_s(r)=-\frac{1}{2}\frac{\mathcal{W}_s N_s}{r^{\frac{(d-1)(1+\omega_s)-2\psi\left(\frac{d-1}{d-2}\right)(1+\omega_s)d}{{1-2\psi\left(\frac{d-1}{d-2}\right)(1+\omega_s)}}}},
\end{eqnarray}
where
\begin{eqnarray}
\mathcal{W}_s=\frac{\left[(d-1)\omega_s-2\psi\left(\frac{d-1}{d-2}\right)(1+\omega_s)\right]\left[2\psi d-(d-2)\right]}{\left[1-2\psi\left(\frac{d-1}{d-2}\right)(1+\omega_s)\right]^2},
\end{eqnarray}
The spacetime metric in (\ref{metric}) depends not only on the dimension $d$ but also on the state parameter $\omega_s$ of the quintessence field. In the case when $d=4$ and $1/l^2=0$, the metric solution in (\ref{metricfunads}) reproduces the results found in \cite{Heydarzade:2017wxu}. Furthermore,  when $\psi=0$, the metric function in Eq.~(\ref{metricfunads}) reduces to the $d$-dimensional AdS black holes in a quintessence background. In addition, for $1/l^2=0$, this also reproduces the $d$-dimensional quintessence black holes \cite{Chen:2008ra}. In the limit where $\omega_s=-1$ and $1/l^2=0$, the spacetime described by Eq.~(\ref{metric}) for $N_s>0$ becomes, 
\begin{eqnarray}\label{omega-1}
ds^2=-\left[1-\frac{m}{r^{d-3}}-N_s r^2\right]dt^2+\frac{dr^2}{\left[1-\frac{m}{r^{d-3}}-N_s r^2\right]}+r^2d\Omega_{d-2}^2.
\end{eqnarray} 
This corresponds to the $d$-dimensional Schwarzschild-Tangherlini-de Sitter black holes \cite{Tangherlini:1963bw}. Interestingly, Eq.~(\ref{omega-1}) does not contain the factor $\psi$ indicating that the Rastall and Einstein theories merge for $\omega_s=-1$. For $N_s<0$ and $\omega_s=-1$, the metric (\ref{metric}) takes the form 
\begin{eqnarray}
ds^2=-\left[1-\frac{m}{r^{d-3}}-\Lambda_{eff} r^2\right]dt^2+\frac{dr^2}{\left[1-\frac{m}{r^{d-3}}-\Lambda_{eff}r^2\right]}+r^2d\Omega_{d-2}^2,
\end{eqnarray}
where $\Lambda_{eff}=(|N_s|+1/l^2)$ represents an effective positive cosmological constant. Consequently, the metric reduces to the $d$-dimensional Schwarzschild-Tangherlini-dS black holes. The metric function (\ref{metricfunads}) for $\xi=2(d-3)\;\text{and}\;N_s=-Q^2$ reduces to that of Reissner-Nordstr$\ddot{o}$m-AdS black hole. As we demonstrate later, the Reissner-Nordstr$\ddot{o}$m-AdS black hole can be viewed as a solution pertaining to the anisotropic quintessence field which can be decomposed as the sum of a perfect fluid and electromagnetic field components. This demonstrats the necessary anisotropy within the quintessence spacetime. In the following subsections, we analyze the dust, radiation, and quintessence fields as subclasses of the solution presented in Eq.~(\ref{metricfunads}). 

As $\psi\to 0$, the metric (\ref{metric}) with the metric function (\ref{metricfunads}) reduces to the $d$-dimensional Kiselev solutions \cite{Kiselev:2002dx} given by
\begin{eqnarray}
\label{kiselev}
ds^2=-\left(1-\frac{m}{r^{d-3}}-\frac{N_s}{r^{(d-3)+(d-1)\omega_s}}\right)dt^2+\left(1-\frac{m}{r^{d-3}}-\frac{N_s}{r^{(d-3)+(d-1)\omega_s}}\right)^{-1}+r^2d\Omega_{d-2}^2.
\end{eqnarray} 
Our objective is to identify the $d$-dimensional black holes within the framework of Rastall theory. As a special case, four-dimensional black hole solutions surrounded by a quintessence field have been thoroughly described in \cite{Heydarzade:2017wxu}, detailing various cases such as dust, radiation, quintessence, and a cosmological constant. The authors of this paper also derived an effective equation of state denoted as $\omega_\text{eff}$, for these different scenarios. We intend to follow a similar guideline to analyze such cases within a generic $d$-dimensional spacetime metric. In the next subsections, we explore these cases in more detail. 
\subsection{Black hole surrounded by a dust field}
As a first limiting case, we consider a dust field surrounding the black hole spacetime. The dust field corresponds to the case where $\omega_s=0$. Therefore, the black hole spacetime described by the metric in Eq.~(\ref{metric}) with the metric function from Eq.~(\ref{metricfunads}) takes the form
\begin{eqnarray}
\label{dust}
ds^2&=&-\left(1-\frac{m}{r^{d-3}}-\frac{N_d}{r^{\frac{(d-3)-2\psi(d-1)}{1-2\psi\left(\frac{d-1}{d-2}\right)}}}\right)dt^2+\left(1-\frac{m}{r^{d-3}}-\frac{N_d}{r^{\frac{(d-3)-2\psi(d-1)}{1-2\psi\left(\frac{d-1}{d-2}\right)}}}\right)^{-1}dr^2\nonumber\\
&+&r^2d\Omega_{d-2}^2,
\end{eqnarray}
where $N_d$ is the dust parameter. It is to be mentioned that if we switch off the Rastall coupling parameter, implying $\psi=0$, we obtain an AdS black hole spacetime with an effective mass given by $m_\text{eff}=m+N_d$. In this case, the black hole spacetime reduces to $$
ds^2=-\left(1-\frac{m_\text{eff}}{r^{d-3}}\right)dt^2+\left(1-\frac{m_\text{eff}}{r^{d-3}}\right)^{-1}dr^2
+r^2d\Omega_{d-2}^2.$$ 
The above spacetime corresponds to a solution in general relativity in the presence of a dust field. If we restore the Rastall theory ($\psi\neq 0$), then our spacetime would correspond to Eq.~(\ref{dust}). Hence, the Rastall parameter plays a crucial role in distinguishing solutions from those in general relativity. \\
We find an effective equation of state parameter $\omega_{eff}$ when we compare Eq.~(\ref{dust}) with Eq.~(\ref{kiselev}). Therefore, the resulting equation of state parameter from the geometry of the Rastall theory is obtained as
\begin{eqnarray}
\label{dusteff}
\omega_{eff}=\frac{1}{d-1}\left(-(d-3)+\frac{(d-3)-2\psi(d-1)}{1-2\psi\left(\frac{d-1}{d-2}\right)}\right).
\end{eqnarray}
One can speculate that $\omega_{eff}$ can never be zero except when $\psi=0$. For $\frac{d-3}{2(d-1)}\leq\psi<\frac{d-2}{2(d-1)}$, we have $\omega_{eff}\leq -(d-3)/(d-1)$.  However, for $\psi<\frac{d-3}{2(d-1)}\cup\psi>\frac{d-2}{2(d-1)}$, we have $\omega_{eff}\geq -(d-3)/(d-1)$, which accounts for the attractive gravitational effect in relation to the strong energy condition.
\subsection{Black hole surrounded by the radiation field}
Next, we consider a radiation field as a second limiting case. The radiation field corresponds to the case where the state parameter $\omega_s=\frac{d-3}{d-1}$. Therefore, the black hole spacetime that accounts for the metric in Eq.~(\ref{metric}) with metric function in Eq.~(\ref{metricfunads}) takes the form
\begin{eqnarray}
\label{rad_1}
ds^2&=&-\left(1-\frac{m}{r^{d-3}}-\frac{N_r}{r^{\frac{2(d-3)-4\psi(d-2)}{1-4\psi}}}\right)dt^2+\left(1-\frac{m}{r^{d-3}}-\frac{N_r}{r^{\frac{2(d-3)-4\psi(d-2)}{1-4\psi}}}\right)^{-1}dr^2\nonumber\\
&+&r^2d\Omega_{d-2}^2,
\end{eqnarray}
where $N_r$ is the radiation field parameter. If we ignore the effect of the Rastall parameter by substituting $\psi=0$, we get spacetime metric (\ref{rad_1}) to have the following form
\begin{eqnarray}
\label{rad_1}
ds^2&=&-\left(1-\frac{m}{r^{d-3}}-\frac{N_r}{r^{2(d-3)}}\right)dt^2+\left(1-\frac{m}{r^{d-3}}-\frac{N_r}{r^{2(d-3)}}\right)^{-1}dr^2+r^2d\Omega_{d-2}^2,
\end{eqnarray}
When $N_r=-Q^2$, the above spacetime reduces to the metric of a $d$-dimensional Reissner-N$\ddot{o}$rdstrom AdS black hole. \\
Moreover, the effective equation of state parameter $\omega_{eff}$ could be determined when we compare Eq.~(\ref{rad_1}) with Eq.~(\ref{kiselev}). Consequently, the equation of state parameter turns out to be
\begin{eqnarray}
\label{dusteff}
\omega_{eff}=\frac{1}{d-1}\left(-(d-3)+\frac{2(d-3)-4\psi(d-2)}{1-4\psi}\right).
\end{eqnarray}
It is realized that $\omega_{eff}$ assumes the value $\omega_{eff}=\frac{d-3}{d-1}$ when $\psi=0$. Therefore, for the range $\frac{d-3}{2(d-2)}\leq\psi<\frac{1}{4}$, we note that $\omega_{eff}\leq \frac{d-3}{d-1}$. On the contrary, for $\psi<\frac{d-3}{2(d-1)}\cup\psi>\frac{d-2}{2(d-1)}$, we have $\omega_{eff}\geq \frac{d-3}{d-1}$. This comparative analysis sharpens our understanding of the interplay between the effective equation of state parameter and the metric function, thereby paving the way for a deeper insight into the more robust features of the radiation fields surrounding black hole spacetimes. 
\subsection{The surrounding quintessence matter field}
As a next example, we discuss the quintessence matter field as a special case when $\omega=-\frac {d-2}{d-1}$. Therefore, the black hole spacetime metric Eq.~(\ref{metric}) with the metric function in Eq.~(\ref{metricfunads}) reduces to the form
\begin{eqnarray}
\label{rad_1}
ds^2&=&-\left(1-\frac{m}{r^{d-3}}-\frac{N_q}{r^{-{\frac{(d-2)(1+2\psi)}{(d-2)-2\psi}}}}\right)dt^2+\left(1-\frac{m}{r^{d-3}}-\frac{N_q}{r^{-{\frac{(d-2)(1+2\psi)}{(d-2)-2\psi}}}}\right)^{-1}dr^2+r^2d\Omega_{d-2}^2,
\end{eqnarray}
where $N_q$ is the parameter of the surrounding quintessence field. The presence of the Rastall parameter ($\psi\neq0$) distinguishes the spacetime qualitatively from the corresponding solutions in Einstein's general relativity. \\
Likewise, we have the effective equation of the state parameter $\omega_{eff}$ when we compare Eq.~(\ref{rad_1}) with Eq.~(\ref{kiselev}). Therefore, the effective equation of the state parameter modifies to give
\begin{eqnarray}
\label{dusteff}
\omega_{eff}=\frac{1}{d-1}\left(-(d-3)-{\frac{(d-2)(1+2\psi)}{(d-2)-2\psi}}\right).
\end{eqnarray}
We ca see that $\omega_{eff}$ can have the form $\omega_{eff}=-\frac{d-2}{d-1}$ when $\psi=0$. Within the range $-\frac{1}{2}\leq\psi<\frac{d-2}{2}$, we have $\omega_{eff}\leq -\frac{d-2}{d-1}$. Furthermore, when $\psi<-\frac{1}{2}\cup\psi>\frac{d-2}{2}$, we have $\omega_{eff}\geq -\frac{d-2}{d-1}$. 

\section{Decomposition of total energy-momentum tensor for the solution (\ref{metric}) with (\ref{metricfunads})}\label{decomp}
In the present section, we discuss the total energy-momentum tensor in the framework of orthonormal tetrads. The basis for the $d$-dimensional spherically symmetric spacetime, as given in Eq~(\ref{metric}), is expressed as \[e^{(a)}_{\mu}=\text{diag}\left(\sqrt{f(r)},1/\sqrt{f(r)},r,r\sin\theta,r\sin\theta\sin\phi,...\right),\] confirming that $g_{\mu\nu}e^{\mu}_{(a)}e^{\nu}_{(b)}=\eta_{(a)(b)},$ where $\eta_{(a)(b)}=\text{diag}(-1,1,...,1)$. Therefore, the total energy-momentum tensor is written as
\begin{eqnarray}
\label{stress-energy}
T^{(a)(b)}_{total}=e^{(a)}_\mu\;e^{(b)}_\nu G^{\mu\nu},\;\;\; G_{(a)(b)}=e^{\mu}_{(a)}e^{\nu}_{(b)}T_{\mu\nu}. 
\end{eqnarray}
As established in \cite{Boonserm:2015aqa, Boonserm:2019phw}, the total energy-momentum tensor $T^{(a)(b)}_{total}$ can be decomposed into the following component forms
\begin{eqnarray}
T^{(a)(b)}_{total}=T^{(a)(b)}_{f}+T^{(a)(b)}_{em}+T^{(a)(b)}_{s},
\end{eqnarray}
where $T^{(a)(b)}_{f}\; T^{(a)(b)}_{em}\;\text{and}\; T^{(a)(b)}_{s}$ represent the energy-momentum tensors for the quintessence field, the electromagnetic field, and the massless minimally coupled scalar field, respectively. These are written as follows.
\begin{eqnarray}
T^{(a)(b)}_{total} &=& \text{diag}[\rho,p_r,p_t...p_t];\;\;T^{(a)(b)}_{f}= \text{diag}[\rho_f,p_f,p_f...p_f];\nonumber\\
T^{(a)(b)}_{em} &=& \frac{E^2}{2}\text{diag}[+1,-1,+1...+1]\;\text{and}\;T^{(a)(b)}_{s}=\frac{(\nabla\phi)^2}{2} \text{diag}[+1,+1,-1...-1],
\end{eqnarray}
where the perfect-fluid parameters in $T^{(a)(b)}_{f} $ are given by
\begin{eqnarray}
p_f=\frac{\left(p_r+p_t\right)}{2},\;\;\;\;\rho_f=\rho-\frac{|p_r-p_t|}{2}.
\end{eqnarray}
The parameters for the electromagnetic and scalar fields are written as
\begin{eqnarray}
E^2=\text{max}\left\lbrace p_t-p_r,0\right\rbrace,\;\;\;\left(\nabla\phi\right)^2=\text{max}\left\lbrace p_r-p_t,0\right\rbrace.
\end{eqnarray} 
We choose the radial coordinate $r$ so that we can focus on the conjugal contributions from the perfect fluid and the electromagnetic fields. For calculation purposes, we explain the metric for the spherically symmetric case without a cosmological constant. Therefore, the metric (\ref{metric}) with its metric function in (\ref{metricfunads}) when $1/l^2=0$ can be written as
\begin{eqnarray}
\label{tetradsmertic}
ds^2=-\left(1-\frac{m}{r^{d-3}}-\frac{N_s}{r^\xi}\right)dt^2+\left(1-\frac{m}{r^{d-3}}-\frac{N_s}{r^\xi}\right)^{-1}dr^2+r^2 d\Omega_{d-2}^2.
\end{eqnarray}
The components of Einstein's tensor in the orthonormal basis are given by
\begin{eqnarray}
\label{stress-energy1}
G{_{{(t)}{(t)}}} &=&-G{_{{(r)}{(r)}}}=-\frac{(d-2)N_s}{2r^{\xi+2}}\left(\xi-(d-3)\right),\nonumber\\
G{_{{(\theta_1)}{(\theta_1})}}&=&G{_{{(\theta_2)}{(\theta_2)}}}=...=G{_{{(\theta_{d-2})}{(\theta_{d-2})}}}=-\frac{(d-2)N_s}{4r^{\xi+2}}\left(\xi^2-(2d-7)\xi+(d-3)(d-4)\right).
\end{eqnarray}
Consequently, the energy-momentum tensors can be written as
\begin{eqnarray}
\label{rho_p}
\rho&=&-p_r=-\frac{(d-2)N_s}{16\pi r^{\xi+2}}\left(\xi-(d-3)\right),\nonumber\\
p_t&=&p{_{\theta_1}}=p_{\theta_2}=...=p_{\theta_{d-2}}=-\frac{(d-2)N_s}{32\pi r^{\xi+2}}\left(\xi^2-(2d-7)\xi+(d-3)(d-4)\right).
\end{eqnarray} 
Next, we analyze the components of the energy-momentum tensor in the context of energy conditions, particularly the null-energy condition. Subsequently, we examine the following equations.
\begin{eqnarray}
\label{NEC1}
p_r-p_t=-(p_t-p_r)=\frac{(d-2)N_s}{32\pi r^{\xi+2}}\left(\xi^2-(2d-9)\xi+(d-3)(d-6)\right),\nonumber\\
\rho+p_r=0,\;\;\rho+p_t=-\frac{(d-2)N_s}{32\pi r^{\xi+2}}\left(\xi^2-(2d-9)\xi+(d-3)(d-6)\right).
\end{eqnarray} 
We note that for the relation $p_r-p_t$, we can have either $(\nabla\phi)^2=0$ or $E=0$, depending on the sign of the quantity $N_s\left(\xi^2-(2d-9)\xi+(d-3)(d-6)\right)$. From Eq.~(\ref{NEC1}), we find that the sign of the quantity $N_s\left(\xi^2-(2d-9)\xi+(d-3)(d-6)\right)$ determines whether the null energy condition is violated or not. We assume the null energy condition is satisfied with the condition $N_s\left(\xi^2-(2d-9)\xi+(d-3)(d-6)\right)<0$. Under this assumption, $p_r-p_t<0$ and hence $(\nabla\phi)^2=0$. With these parameters, the energy-momentum tensors for the perfect fluid are expressed as
\begin{eqnarray}
p_f&=&\frac{(p_r+p_t)}{2}=\frac{N_s}{32\pi r^{\xi+2}}\left(-\xi^2+3(d-3)-2(d-3)^2\right),\nonumber\\
\rho_f&=&=\rho-\frac{|p_r-p_t|}{2}=-\frac{N_s}{32\pi r^{\xi+2}}\left(-\xi^2+3(d-3)-2(d-3)^2\right),\nonumber\\
w_f&=&\frac{p_f}{\rho_f}=-1.
\end{eqnarray}
The contribution of electromagnetic part, $T^{(a)(b)}_{em}$, from the total electromagnetic tensor, we evaluate
\begin{eqnarray}
E^2=\frac{N_s}{16\pi r^{\xi+2}}\left(-\xi^2+(d-5)\xi+2(d-3)\right).
\end{eqnarray}
Consequently, we can write the energy-momentum tensor of the perfect fluid as
\begin{eqnarray}
T^{(a)(b)}_{f}&=&\frac{N_s}{32\pi r^{\xi+2}}\left(-\xi^2+3(d-3)-2(d-3)^2\right)\text{diag}(-1,+1,+1,...,+1),\nonumber\\
T^{(a)(b)}_{f}&=&\frac{N_s}{16\pi r^{\xi+2}}\left(-\xi^2+(d-5)\xi+2(d-3)\right)\text{diag}(+1,-1,+1,...,+1),
\end{eqnarray}
and hence we can express the linear decomposition of the total energy-momentum tensor as
\begin{eqnarray}
T^{(a)(b)}_{total}&=&T^{(a)(b)}_{f}+T^{(a)(b)}_{em}=-\frac{(d-2)N_s}{16\pi r^{\xi+2}}\left(\xi-(d-3)\right)\times\nonumber\\
&&\text{diag}\left(+1,-1,\frac{\xi^2-(2d-7)\xi+(d-3)(d-4)}{(d-2)(\xi-(d-3))},...\frac{\xi^2-(2d-7)\xi+(d-3)(d-4)}{(d-2)(\xi-(d-3))}\right).\nonumber\\
\end{eqnarray}
Next, we check the validity of the electric field strength within the constraints of the fluid state parameter, $\omega_s$, and the quintessence parameter, $N_s$. The field strength tensor yields
\begin{eqnarray}
E(r)=\pm \sqrt{\frac{{|(\xi^2-(d-5)\xi-2(d-3))N_s|}}{16\pi}}\frac{1}{r^{\xi/2+1}},\;\;\;\;\frac{dE}{dr}=-\frac{E(\xi+2)}{2r}. 
\end{eqnarray}
Invoking Gauss's law, we derive the charge enclosed within a sphere of radius $r$,
\begin{eqnarray}
Q(r)=E(r)\Omega_{d-2}r^{d-2}=\pm \Omega_{d-2}\sqrt{\frac{{|(\xi^2-(d-5)\xi-2(d-3))N_s|}}{16\pi}}\frac{1}{r^{(\xi-2(d-3))/2}}.
\end{eqnarray}
The electric charge density subsequently resulted into the following expression
\begin{eqnarray}
\sigma_{em}=\frac{dQ}{dV_{d-1}}=\frac{1}{\Omega_{d-2}r^{d-2}\sqrt{g_{rr}}}\frac{dQ}{dr}=\sqrt{g^{rr}}\left(\frac{dE}{dr}+\frac{(d-2)E}{r}\right)=\sqrt{g^{rr}}\left(\frac{E(-\xi+2(d-3)}{2r}\right).\nonumber\\
\end{eqnarray}
Inserting our previous expression for $E(r)$, we observe that the charge density is reformulated to be
\begin{eqnarray}
\sigma_{em}=\pm \sqrt{g^{rr}} \sqrt{\frac{{|(\xi^2-(d-5)\xi-2(d-3))N_s|}}{16\pi}}\frac{(2(d-3)-\xi)}{2r^{\xi/2+2}}.
\end{eqnarray}
Notably, when $\xi=2(d-3)$, the charge density diminishes to zero, thereby simplifying our metric function for $N_s=-Q^2$, and hence the resulting metric reduces to the Reissner-N$\ddot{o}$rdstrom spacetime, wherein all charges are concentrated at the origin. Under this scenario, the equation of state parameter for $\xi=2(d-3)$ is evaluated to be
\begin{eqnarray}
\omega_s=\frac{(d-3)(d-2)-2\psi (d-1)(d-4)}{(d-1)((d-2)+2\psi(d-4))}.
\end{eqnarray}
Obviously, when $\psi=0$, the quantity $\omega_s$ reduces to $w_s=(d-3)/(d-1)$ \cite{Chen:2008ra}. With this machinery in hand, our analysis of the linear decomposition of the total energy-momentum tensor leads us to a compelling conclusion--irrespective of the circumstances, our results must be accompanied by the linear combinations of the energy-momentum tensor of the perfect fluid and the electromagnetic field components, provided the null energy condition remains valid. 
\section{Horizon structure}
\label{horizon_struc}
\begin{figure}
\includegraphics[scale=0.62]{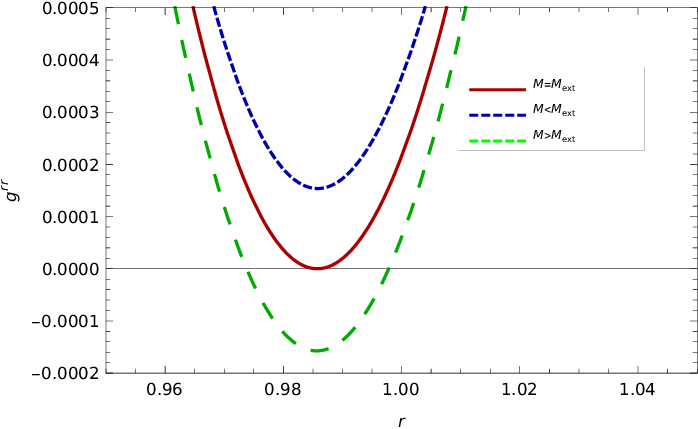}\hspace{2mm}
\includegraphics[scale=0.62]{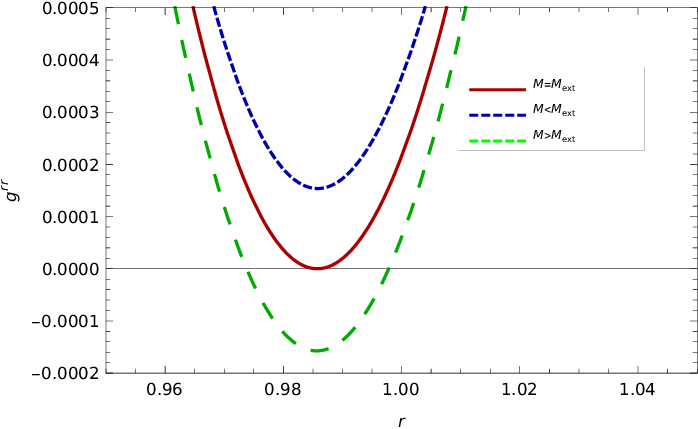}\\
\includegraphics[scale=0.65]{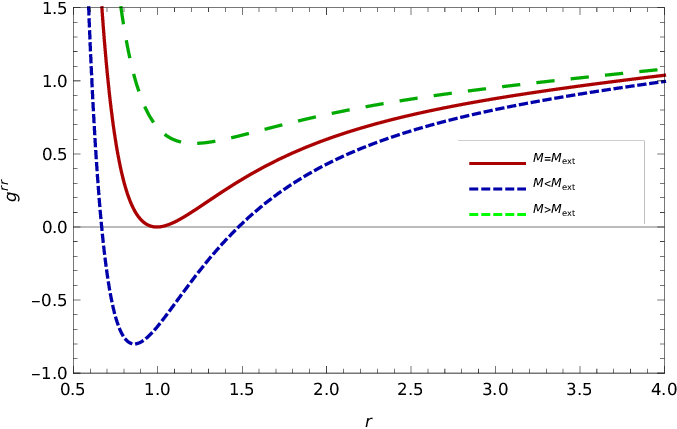}\hspace{2mm}
\includegraphics[scale=0.65]{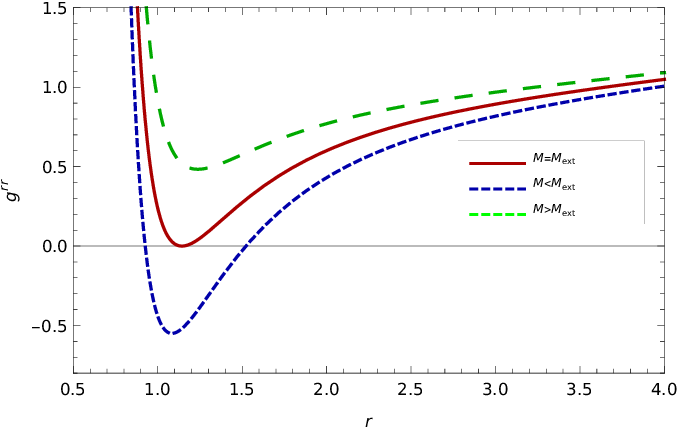}\\
\includegraphics[scale=0.65]{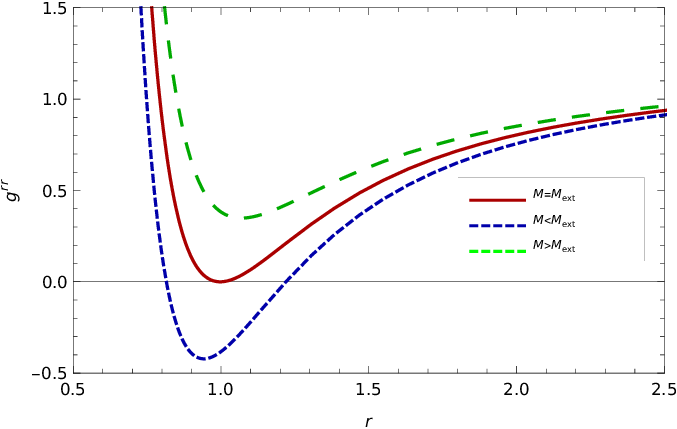}
\includegraphics[scale=0.65]{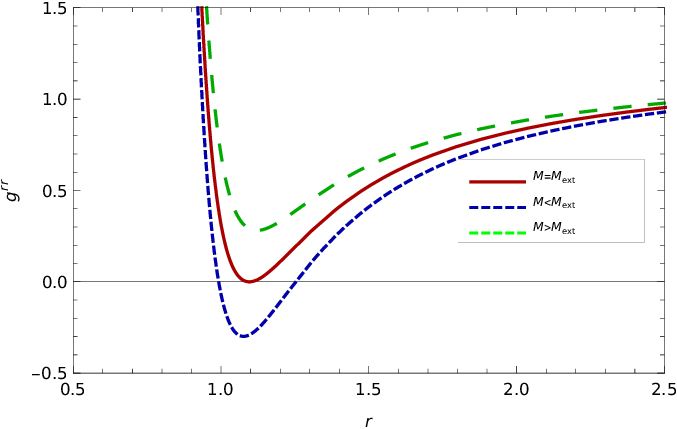}
 \caption{Plots showing the behavior of the metric function ($g^{rr}$) as a function of radial coordinate ($r$) for the charged black hole in the AdS spacetime in Rastall gravity. \label{fighor}}
\end{figure}

To understand the horizon structures, we must identify the radial coordinate at which a coordinate singularity appears. It is mathematically established that the radius of the horizon occurs at the point where $g^{rr}=f(r)=0$. Solving this equation reveals the possible existence of the Cauchy (inner) and black hole event (outer) horizons. However, obtaining an analytical expression for the horizon radii is not feasible due to the complexity of the polynomial equation involving higher powers of the radial coordinate $r$. Therefore, we solve this equation numerically for specific values of the Rastall coupling parameter ($\psi$), spacetime dimensions $d$, parameter $N_s$, and the AdS curvature radius $l$. Fig.~\ref{fighor} illustrates three possible configurations of the horizon structure. In each plot, when $M=M_{\text{ext}}$, where $M_\text{ext}$ is the mass of an extremal configuration, we observe two coinciding horizons. This occurs in extremal configurations when the Cauchy and black hole event horizons coincide. For values of $M>M_{\text{ext}}$, there are no horizons, meaning that black hole configurations do not exist. Two distinct horizons are present when $M<M_{\text{ext}}$. The condition for extremality is met when $$f(r_h)=0=\left.\frac{\partial f}{\partial r}\right\vert_{r=r_h}.$$.  By solving these equations, we can confidently determine the extremal horizon radius and mass of the corresponding extremal black hole configurations. To analyze the horizon properties, we set $\psi=(0,0.2)$, and $N_s=-Q^2$, with $Q=1$ in each spacetime dimension. We focus solely on the charged case, and we limit our analysis to these specific values for the parameters $\psi$ and $Q$. The figures in the left column correspond to the case where $\psi=0$, while the right column represents $\psi=0.2$.  Furthermore, in each column, we set the spacetime dimensions starting from $d=4$ at the top and proceeding to $d=6$ at the bottom. Throughout our analysis we restrict ourselves to these specific set of values of the Rastall coupling parameter $\psi$ and the spacetime dimension $d$. 
\section{Thermodynamics}\label{thermodynamics}
In this section, we discuss the thermodynamics of $d$-dimensional AdS black holes surrounded by a quintessence field. The metric element (\ref{metric}) is invariant under the transformation $t\to -t$, and correspondingly, we have a time-like Killing vector of the form $\xi^{\mu}=\delta{^\mu}{_t}$. The Killing vector $\xi^{\mu}$ is a null generator of the event horizon, satisfying $\xi{^\mu}\xi{_\mu}=0$. Consequently, we obtain the relation $g_{tt}\vert_{r=r_h}=g^{rr}\vert_{r=r_h}=f(r_h)=0$. This condition is used to determine the radius of the event horizon, which has a complex structure that cannot be solved analytically. The ADM mass $M$ is obtained by solving $f(r_h)=0$ in terms of the event horizon radius $r=r_h$. It is given by the expression
\begin{eqnarray}\label{ADM1}
M&=&\frac{(d-2)\Omega_{d-2}}{16\pi}\Bigg[1-\frac{N_s}{r_h^{\xi}}+\frac{r_h^2}{l^2}\Bigg]{r_h^{d-3}}.
\end{eqnarray} 
The surface gravity is defined as $\kappa=\sqrt{-\nabla_\mu\xi_{\nu}\nabla^\mu\xi^{\nu}/2}$ and is related to the temperature $T$ through $T=\kappa/2\pi$. Thus, the temperature of the black holes in $d$-dimensional spacetime is expressed as
\begin{eqnarray}\label{tempads}
T&=&\frac{f^{\prime}(r_h)}{4\pi}=\frac{1}{4\pi}\Bigg[\frac{(d-3)}{r_h}+(d-1)\frac{r_h}{l^2}+{(3-d+\xi)}\frac{N_s}{r_h^{1+\xi}}\Bigg].
\end{eqnarray}
The entropy can be calculated as follows.
\begin{eqnarray}\label{entropy}
S=\int T^{-1}\left(\frac{\partial M}{\partial r_h}\right)dr_h=\frac{\Omega{_{d-2}}r_h^{d-2}}{4}.
\end{eqnarray}
In extended phase space of black hole thermodynamics, the cosmological constant can be identified as thermodynamic pressure. For the metric given in Eq.~(\ref{metricfunads}), this is written as
\begin{eqnarray}\label{pres}
P&=&-\frac{\Lambda}{8\pi}=\frac{(d-1)(d(1-2\psi)-2)}{16\pi l^2}
\end{eqnarray}
The expression for mass (\ref{ADM1}) in terms of pressure (\ref{pres}) can now be reformulated as
\begin{eqnarray}\label{ADM2}
M&=&\frac{(d-2)\Omega_{d-2}}{16\pi}\Bigg[1-\frac{N_s}{r_h^{\xi}}+\frac{16\pi\;P}{(d-1)(d(1-2\psi)-2)}r_h^2\Bigg]{r_h^{d-3}}
\end{eqnarray}
We observe that the entropy, Eq.~(\ref{entropy}) does not contain the contribution from the quintessence field as well as the Rastall parameter $\psi$. However, the thermodynamic pressure, Eq.~(\ref{pres}) is independent of the parameter $N_s$ but not of the Rastall parameter $\psi$ and hence affects the thermodynamics of the black holes in the extended phase space.\\
It is suggested that the first law of black hole mechanics must include the variation of the cosmological constant as a thermodynamic variable when written for AdS black holes. This treatment leads us to consider the cosmological constant as pressure and its conjugate quantity as thermodynamic volume in the extended phase space of black hole thermodynamics. The thermodynamics in such an extended phase space have been studied in various theories of gravity. To confront the idea of phase transition of the black hole system with the usual classical van der Waals system such treatments have been provided to investigate the critical phenomena and related pathologies \cite{Kubiznak:2012wp, Gunasekaran:2012dq, Altamirano:2013ane, Kastor:2009wy, Altamirano:2014tva, Altamirano:2013uqa, Frassino:2014pha, Wei:2014hba, Dolan:2014vba, Wei:2015ana, Hennigar:2016xwd, Dolan:2011jm, Dolan:2013dga, Chamblin:1999tk, Dolan:2011xt, Mo:2014mba, Banerjee:2011au, Belhaj:2014tga}. Therefore, the first law of black hole thermodynamics in the extended phase space reads
\begin{eqnarray}\label{ex1st}
dM=TdS+VdP+\Theta_s dN_s,
\end{eqnarray}
where $S$ is the entropy and $P$ is related to the cosmological constant of the black holes defined in Eqs.~(\ref{entropy}) and (\ref{pres}). Here we treat $\Theta_s$ as a generalized force conjugate to the surrounding quintessence field structure parameter $N_s$ and are introduced to make the first law consistent with the Smarr-Gibbs-Duhem relation. Thus on using Eqs.~(\ref{ADM2}) and~(\ref{ex1st}), we obtain $\Theta_s$ having the following form \cite{Chen:2008ra}
\begin{eqnarray}
\Theta_s &=&\left(\frac{\partial M}{\partial N_s}\right)_{S,P,Q}=-\frac{(d-2)\left(2\pi\right)^{\frac{d-1}{2}}}{16\pi \Gamma{\frac{d-1}{2}}}\frac{1}{r_h^{\xi}}.
\end{eqnarray}
The thermodynamic volume term from Eq.~(\ref{ex1st}) reads
\begin{eqnarray}\label{vol}
V&=&\left(\frac{\partial M}{\partial P}\right)_{S,Q,N_s}
\end{eqnarray}
Using Eqs.~(\ref{pres}) and (\ref{ADM2}), the Eq.~(\ref{vol}) is obtained as
\begin{eqnarray}\label{vol1}
V=\frac{(d-2)\Omega_{d-2}r_h^{d-1}}{(d-1)(d(1-2\psi)-2)}
\end{eqnarray}
Therefore, the corresponding Smarr-Gibbs-Duhem formula \cite{Kastor:2009wy,Altamirano:2014tva} when using the Euler's theorem is obtained as
\begin{eqnarray}\label{smarr}
\frac{d-3}{d-2}M&=&TS-\frac{2}{d-2}VP+\frac{\xi}{d-2}\Theta_s N_s.
\end{eqnarray}
The mass depends on the state parameter $\omega_s$ of the quintessence field as it is contained in $\xi$. In the limit $\omega_s\to -1$, the last term in the right hand side of Eq.~(\ref{smarr}) becomes $-\frac{2}{d-2}\;\Theta_s N_s$. Setting $N_s=1/l^2$, we have $\Theta_{l}=\frac{\partial M}{\partial l}\frac{\partial l}{\partial N_s}=-\frac{2}{l^3}\Theta_s$ and $-\frac{2}{d-2}\;\Theta_s N_s$=$\frac{1}{d-2}\Theta_l\;l$. The fact that if $\Lambda$ and hence $P$ is treated indeed as a constant, the second last term in (\ref{ex1st}) is essentially zero. However, it is not a vanishing quantity and can have physical consequences. It is a common belief that in the inflationary models, $\Lambda$ can be identified as a variable quantity \cite{Dolan:2011jm, Dolan:2011xt}. Among the black hole research community, it is widely accepted that when analyzing the thermodynamics of extended phase space, the van der Waals theory serves as a phenomenological description of a \textit{real} gas system. However, it starts to break down below the critical temperature. In this region, the isotherms exhibit the oscillatory behavior in the pressure-volume ($P-V$) plane, leading to development of a pair of extremal points. In the subsequent analysis, we observe how the isotherms in the ($P-V$) plane behave below the critical points. 

\section{P-V criticality}\label{PV_criticality}
We explore the critical phenomena of the black holes under thermodynamic equilibrium. The Hawking temperature (\ref{tempads}), 
when expressed in terms of pressure (\ref{pres}), is expressed as
\begin{eqnarray}\label{tempadsp}
T&=&\frac{1}{4\pi}\Bigg[\frac{(d-3)}{r_h}+\frac{16\pi r_h P}{(d(1-2\psi)-2)}+{(3-d+\xi)}\frac{N_s}{r_h^{1+\xi}}\Bigg]
\end{eqnarray}
Before solving this equation for pressure, we need to introduce the specific volume of the van der Waals-like fluid. This allows us to express the equation of state that relates pressure to other thermodynamic quantities such as temperature, volume, and parameters that characterize AdS black holes in arbitrary dimensions, with the inclusion of the quintessence field in Rastall theory, which can be expressed as $P=P(T, V, Q, N_s)$. The fluid volume as discussed in \cite{Gunasekaran:2012dq, Kubiznak:2012wp}, is expressed in the form
\begin{eqnarray}\label{volume1}
v=\frac{4\ell^{d-2}_{P}}{d-2}r_h=\frac{4\ell^{d-2}_{P}}{d-2}\left({\frac{(d-1)(d(1-2\psi)-2)V}{(d-2)\Omega_{d-2}}}\right)^{\frac{1}{d-1}}
\end{eqnarray}
In geometric units, where $\ell_{P}=1$, we have 
\begin{eqnarray}\label{r_h}
r_h=\frac{d-2}{4}v
\end{eqnarray}
Thus, replacing $r_h$ as defined above, we obtain the equation of state from the Hawking temperature in Eq.~(\ref{tempadsp}), as follows
\begin{eqnarray}\label{eosp}
P&=&\frac{(d(1-2\psi)-2)}{d-2}\frac{T}{v}-\frac{(d-3)(d(1-2\psi)-2)}{\pi(d-2)^2 v^2}-\frac{{(3-d+\xi)(d(1-2\psi)-2)N_s}}{16\pi\left(\frac{d-2}{4}\right)^{2+\xi}v^{2+\xi}}
\end{eqnarray}
We define a characteristic temperature, $T_0$, which is also present in the Van der Waals equation. This is given by the formula $$T_0=\frac{(d-3) \xi  \left(\frac{(\xi +1) (d-2)^{-\xi } (d-\xi -3) N_s}{d-3}\right){}^{-1/\xi }}{4 \pi  (d-2) (\xi +1)}.$$ 
Below this temperature, the pressure becomes negative for some $$v_0=4 \left(\frac{(\xi +1) (d-2)^{-\xi } (d-\xi -3) N_s}{d-3}\right){}^{1/\xi }.$$ Similar to the behavior of Van der Waals fluid, the $\tilde{P}-\tilde{v}$ diagram exhibits oscillatory behavior, which is completely unphysical. As we already mentioned, this oscillatory behavior in the isotherms must be replaced with an isobaric plateau. When 

In the limit as $\psi\rightarrow 0$, this equation of state (\ref{eosp}) reduces to describe $d$-dimensional AdS black holes in quintessence background, following Einstein's GR. If both $N_s$ and $\psi$ vanish, the equation of state (\ref{eosp}) simplifies to that of Schwarzschild-Tangherlini-AdS black holes \cite{Gunasekaran:2012dq}. \\
The critical points are determined by the following conditions.
\begin{eqnarray}\label{critical}
\left(\frac{\partial P}{\partial v}\right)_{T=T_c}=\left(\frac{\partial^2 P}{\partial^2 v}\right)_{T=T_c}=0.
\end{eqnarray}
We can solve the above equations (\ref{critical}) analytically. Therefore, the critical quantities now read
\begin{eqnarray}\label{cvol}
v_c&=&\left[-\frac{(3-d+\xi)(1+\xi)(2+\xi)N_s}{2(d-3)}\right]^{\frac{1}{\xi}}
\end{eqnarray}
\begin{eqnarray}\label{ctemp}
T_c&=&\frac{2(d-3)\xi}{\pi (d-2)(1+\xi)v_c}
\end{eqnarray}
\begin{eqnarray}\label{cp}
P_c&=&\frac{(d(1-2\psi)-2)(d-3)\xi}{\pi (d-2)^2(2+\xi)v_c^2}
\end{eqnarray}
The subscript ``c" denotes the values of the physical quantities at the critical points. Eqs.~(\ref{cvol}) and (\ref{ctemp}) together with  Eq.~(\ref{cp}) lead to the following universal ratio.
\begin{eqnarray}\label{universal}
\rho_c=\frac{P_c v_c}{T_c}=\frac{d(1-2\psi)-2}{2(d-2)}\left(\frac{1+\xi}{2+\xi}\right)
\end{eqnarray}
This ratio is independent of the surrounding field parameter $N_s$, but is dependent on the Rastall coupling constant $\psi$ and spacetime dimensions $d$. In Einstein's GR, $\psi\rightarrow 0$, we get the universal ratio for $d$-dimensional AdS black holes in a quintessence background. When $\psi=0$ and $\omega_s=\frac{d-3}{d-1}$, the universal ratio corresponds to the $d$-dimensional Reissner-Nordstr$\ddot{o}$m-AdS black holes ($N_s=-Q^2$) \cite{Gunasekaran:2012dq}, which reads 
\begin{eqnarray}
\rho_c=\frac{2d-5}{4d-8}
\end{eqnarray}
We recover the ratio $\rho_c=3/8$ in $d=4$, a universal feature of the vdW fluid. It is seen that the critical parameters, Eqs.~(\ref{cvol})-(\ref{cp}), depend on the fluid parameter and therefore reflect in this way the presence of some exotic matter fields, e.g., dark energy, on the critical behavior of the vdW fluid. If we put $\psi=0,\;\text{and}\;N_s=-Q^2$, all these critical quantities reduce to those of $d$-dimensional RN-AdS black holes of $d$ dimensions \cite{Gunasekaran:2012dq}.
\begin{figure*}
\includegraphics[scale=0.65]{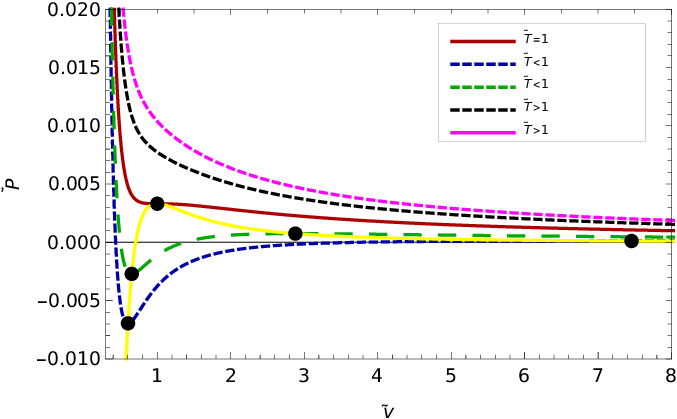}
\includegraphics[scale=0.65]{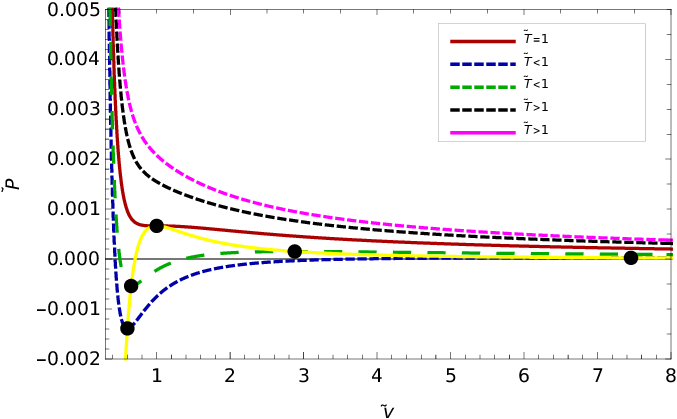}\\
\includegraphics[scale=0.65]{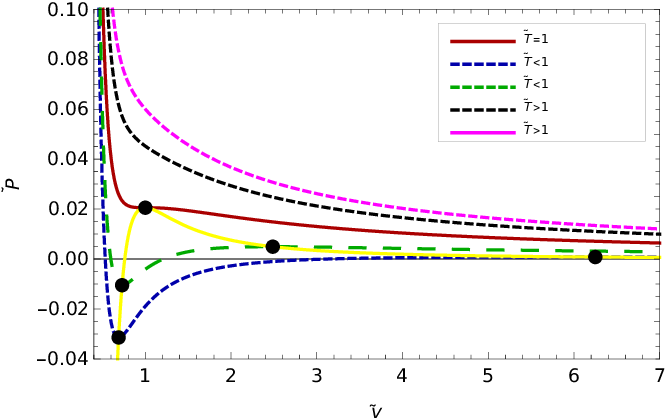}
\includegraphics[scale=0.65]{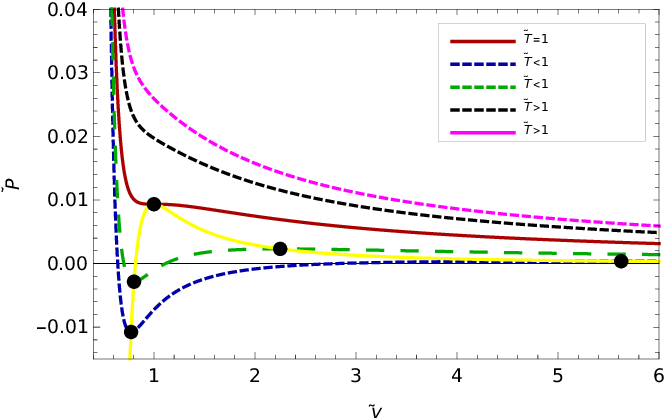}\\
\includegraphics[scale=0.65]{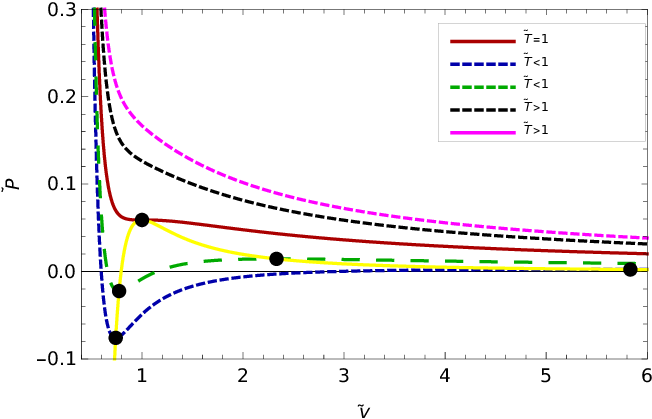}
\includegraphics[scale=0.65]{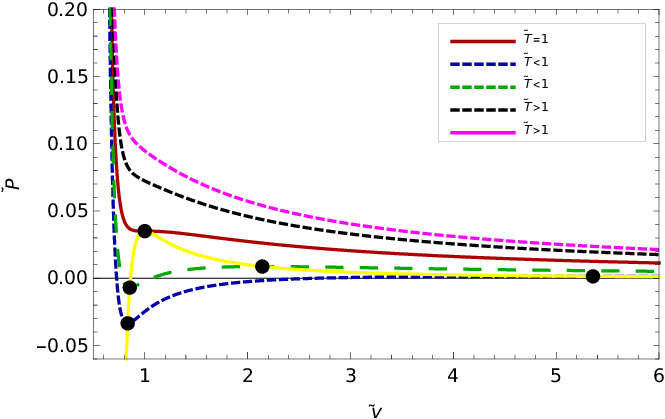}
 \caption{Plots showing the behavior of reduced pressure $\tilde{P}$ versus reduced specific volume $\tilde{v}$ for the charged black hole in the AdS spacetime in Rastall gravity. \label{figPressure}}
\end{figure*}
\begin{figure*}
\includegraphics[scale=0.65]{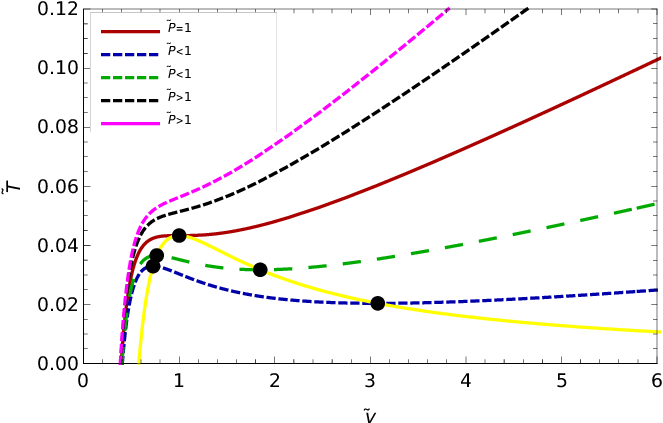}
\includegraphics[scale=0.65]{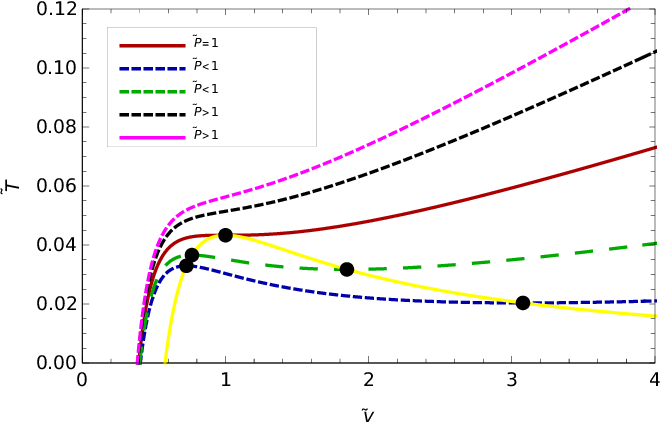}\\
\includegraphics[scale=0.65]{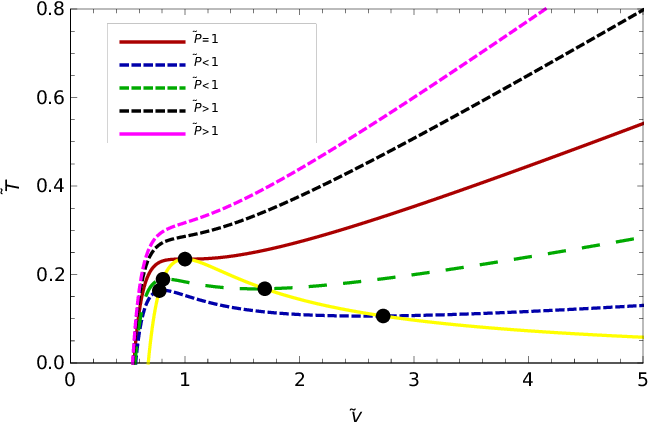}
\includegraphics[scale=0.65]{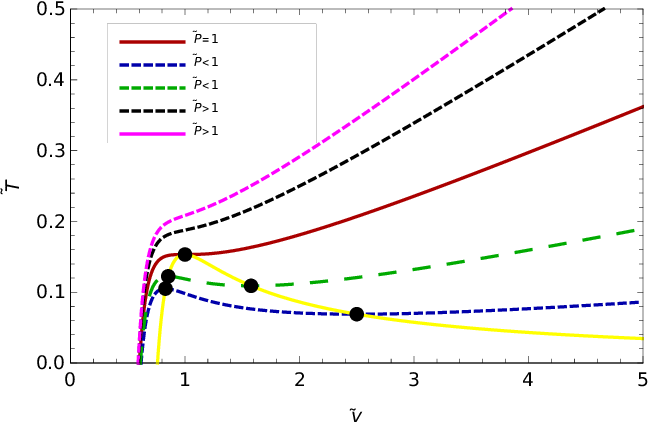}\\
\includegraphics[scale=0.65]{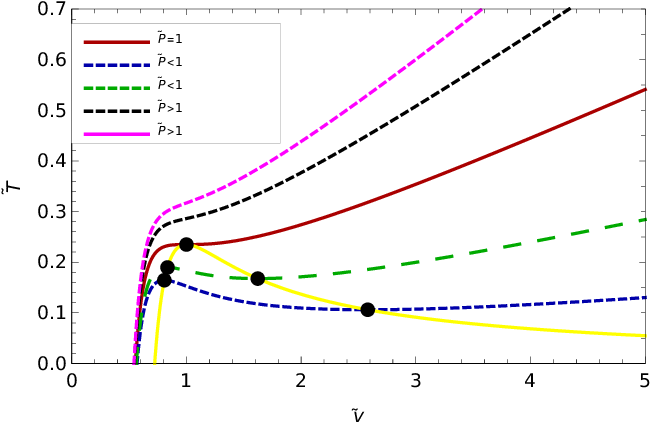}
\includegraphics[scale=0.65]{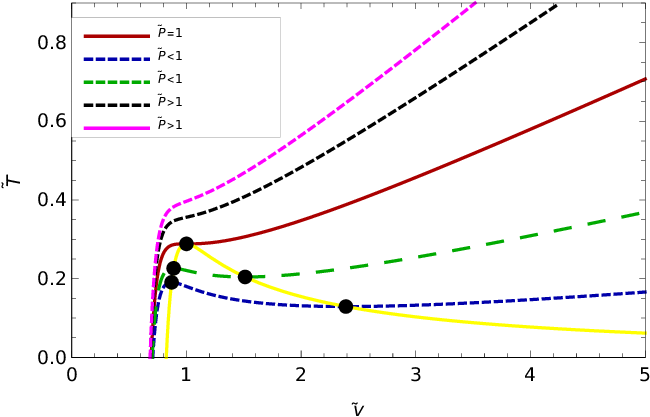}
 \caption{Plots showing the behavior of reduced pressure $\tilde{T}$ versus reduced specific volume $\tilde{v}$ for the charged black hole in the AdS spacetime in Rastall gravity.} 
 \label{figTemp}
\end{figure*}
Before we concentrate on the critical behavior of isotherms, we redefine various thermodynamic quantities in a dimensionless manner
$$\tilde{T}=\frac{T}{T_c},\;\tilde{P}=\frac{P}{P_c},\;\tilde{v}=\frac{v}{v_c},\;\tilde{r}=\frac{r}{r_c},\;\tilde{G}=\frac{G}{G_c}$$ and so on.
Fig.~\ref{figPressure} shows the nature of the $\tilde{P}-\tilde{v}$ diagram for the charged ($N_s=-Q^2$) AdS black hole in the Rastall theory. The temperature of the isotherms decreases as we move from the bottom to the top of the diagram. The upper two curves correspond to the behavior of the ideal one-phase gas, which corresponds to $\tilde{T}>1$. The critical isotherm occurs at $\tilde{T}=1$, while $\tilde{T}<1$ resembles the oscillatory behavior of the van der Waals-like fluid. We also investigated the properties of the isobaric curve shown in Fig.~\ref{figTemp}. Similarly to the $\tilde{P}-\tilde{v}$ diagram, the $\tilde{T}-\tilde{v}$ curve exhibits oscillatory behavior when the reduced pressure is $\tilde{P}<1$. We get the critical isobar for the case where $\tilde{P}=1$, while for $\tilde{P}>1$, we have the monotonically varying isobars. Therefore, the extended phase-space thermodynamics allows us to investigate the AdS black hole phase transitions in the $\tilde{P}-\tilde{v}$ and $\tilde{T}-\tilde{v}$ planes, providing a clear formulation of the vdW-like picture. In this region, the isotherms in the $\tilde{P}-\tilde{v}$ plane and the isobar in the $\tilde{T}-\tilde{v}$ plane start to develop a pair of extremal points. Between the minimum and maximum, an increase in volume leads to an unexpected rise in pressure, which sharply contradicts experimental results that indicate the presence of an isobaric plateau in the $\tilde{P}-\tilde{v}$ plane. This discrepancy between theoretical prediction and experimental data can be reconciled by replacing the oscillating part of the isotherms with a horizontal isobar, in accordance with Maxwell's equal-area construction. The Rastall coupling parameter $\psi$ and the dimension of the spacetime $d$ are expected to significantly influence the nature of isothermal or isobaric curves. In four dimensions, the behavior of the maxima of these curves is almost independent of the Rastall parameter. However, as we increase the spacetime dimension, the maxima in the isothermal and isobaric curves shift towards higher values of the pressure and the temperature, respectively. The yellow dashed curves in Figs.~\ref{figPressure} and \ref{figTemp} indicate the extremal points, which are determined either by $(\partial_v P)_T=0$ or by $(\partial_v T)_P=0$. Then, we can substitute the expressions for $T$ and $P$ in Eqs.~(\ref{eosp}) and (\ref{tempadsp}), respectively. 

\begin{figure}[h]
\includegraphics[scale=0.6]{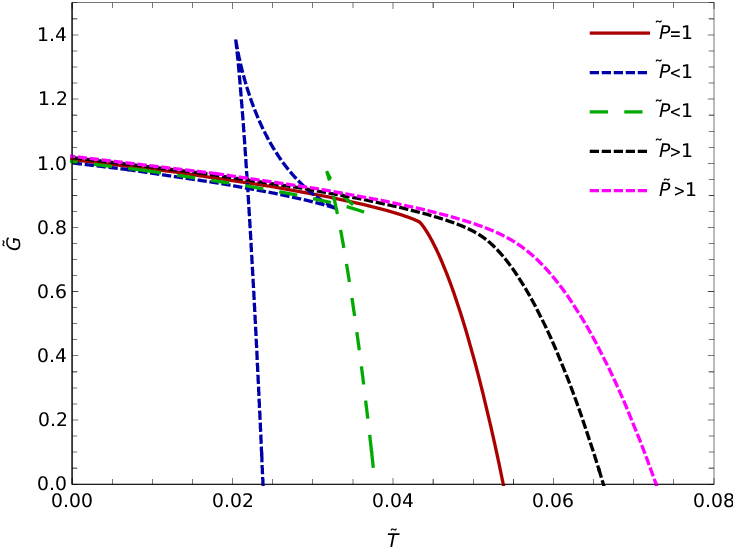}\hspace{2mm}
\includegraphics[scale=0.6]{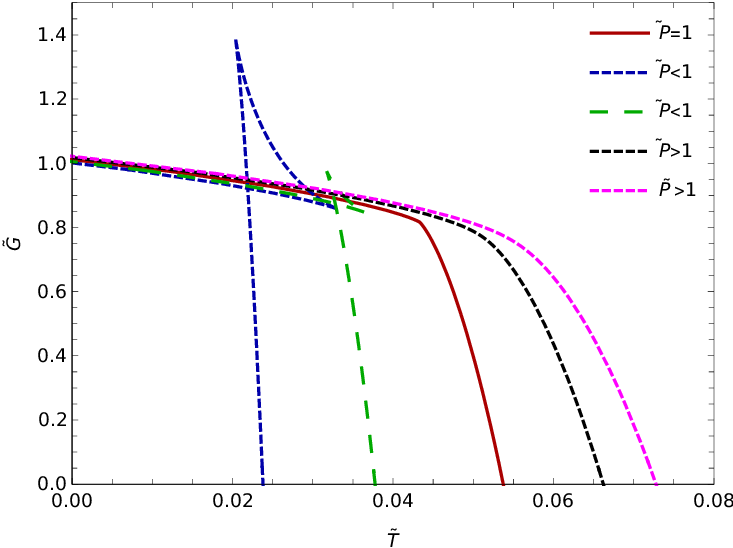}\\
\includegraphics[scale=0.6]{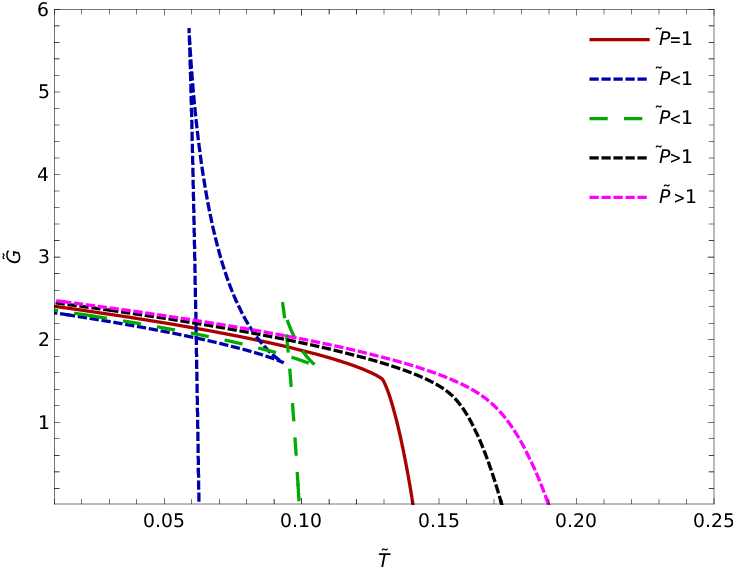}\hspace{2mm}
\includegraphics[scale=0.6]{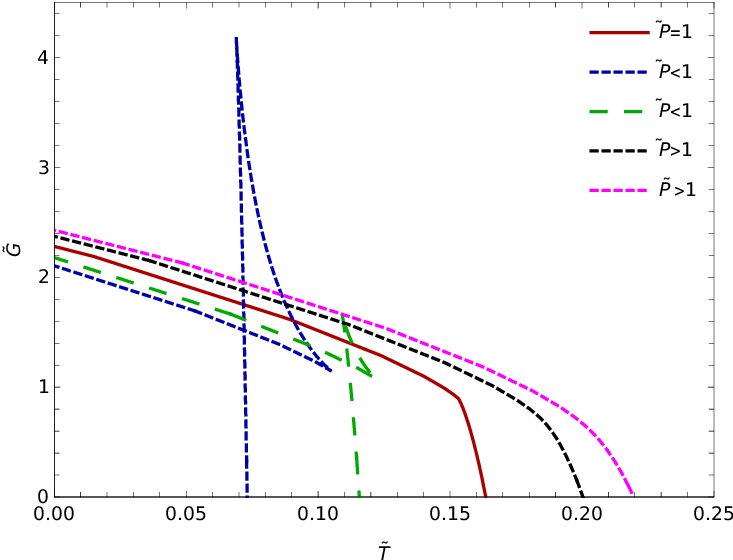}\\
\includegraphics[scale=0.6]{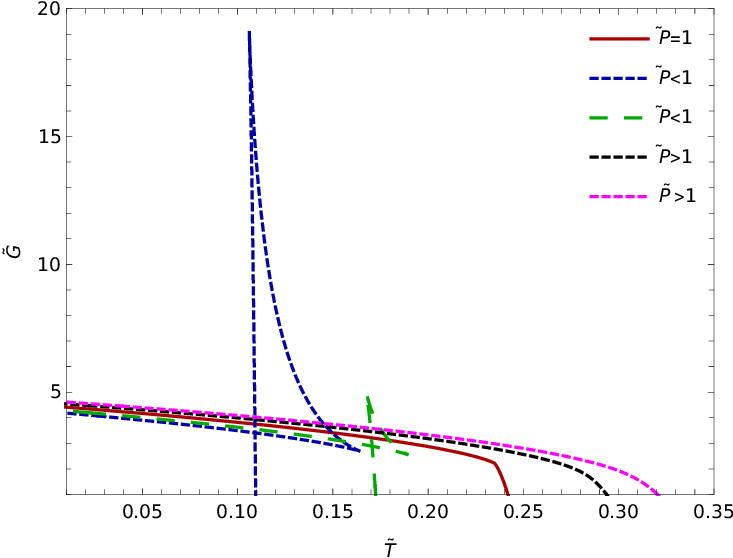}\hspace{2mm}
\includegraphics[scale=0.6]{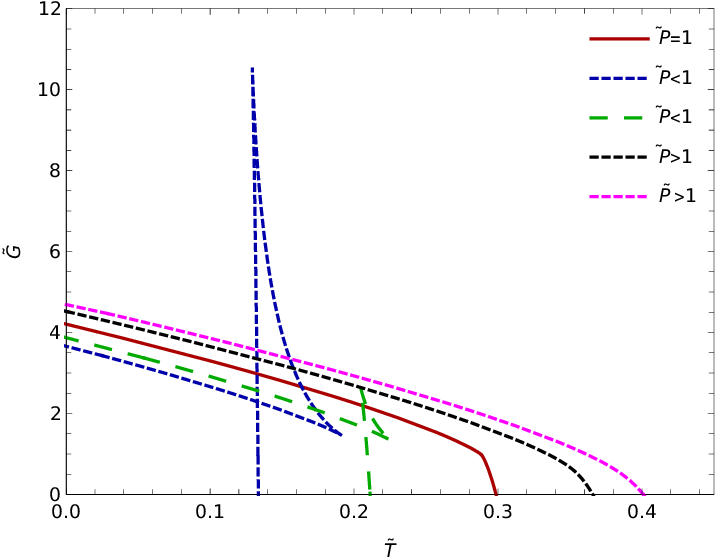}
 \caption{Plots showing the behavior of reduced Gibbs free energy $\tilde{G}$ vs the temperature $\tilde{T}$ for the charged black hole in the AdS spacetime in Rastall gravity. \label{fig_gibbs}}
\end{figure}

Next, we derive another important thermodynamic quantity in equilibrium physics, the Gibbs free energy of the thermodynamic system. In the extended phase space, the enthalpy of the thermodynamic system is interpreted as the mass of the black holes. By treating mass as an enthalpy, we can obtain the Gibbs free energy, $G=H-TS=M-TS$ \cite{Chamblin:1999tk}. This results in the following expression--
\begin{eqnarray}\label{gibbs}
G&=&\frac{\Omega_{d-2}}{16\pi}\Big[r_h^{d-3}-\frac{16\pi r_h^{d-1}P}{(d-1)(d(1-2\psi)-2)}-\frac{(1+\xi)N_s}{r_h^{\xi+3-d}}\Big].
\end{eqnarray} 
We have a thermodynamically preferred branch of an AdS black hole system when the Gibbs energy is minimal for a fixed value of temperature. Phase transitions can occur as $\tilde{G}-\tilde{T}$ intersects when $\tilde{v}$ is varied. Fig.~\ref{fig_gibbs} shows the nature of the Gibbs free energy as a function of temperature for fixed values of the charge $Q$ and the Rastall parameter in various space-time dimensions $d$. The top two curves in Fig.~\ref{fig_gibbs} correspond to the case where $\tilde{P}>1$, while the region where $\tilde{P}<1$ corresponds to the case exhibiting swallow-tail behavior. Notably, for $\tilde{P}<1$, there is a first-order phase transition from small to large black holes. We choose the specific values of the Rastall parameter and the field parameter ($Q=1$). In the left panel of Fig.~\ref{fig_gibbs}, we set $\psi=0$ and $Q=1$, but the spacetime dimension changes from $d=4$ to $d=6$ from top to bottom. Similarly, in the right panel, we set $\psi=0.2$ and $Q=1$. To be more specific, when $\tilde{P}<1$, the $\tilde{G}-\tilde{T}$ diagram shows three distinct characteristics for the black hole states, namely, small stable black hole, unstable intermediate black hole, and large stable black hole. Similarly to that of an ordinary thermodynamic system, we can have an associated order of phase transition in black hole thermodynamics. We define the order of a phase transition as the lowest-order derivative of the Gibbs free energy $\tilde{G}$ undergoing discontinuity at the phase transition point. As illustrated in Fig.~\ref{fig_gibbs}, when the reduced pressure $\tilde{P}<1$, we observe a first-order phase transition where the two distinct branches intersect, while for $\tilde{P}=1$, there is a second-order phase transition. For $\tilde{P}>1$, there exists a single phase without any phase transition.

\subsection{Analytical study of classical Ehrenfest equations}\label{check}
This section focuses on the analytical study of the order of phase transitions at critical points. When the Clausius-Clapeyron equation is completely satisfied, we classify the phase transition as first-order. In contrast, a distinct second-order phase transition occurs at critical points, which can be successfully analyzed using the Ehrenfest equations. These equations have been successfully applied to van der Waals fluids within the context of the extended phase space of black hole thermodynamics, encompassing various black hole solutions in Einstein's General Relativity (GR) and other modified theories of gravity \cite{Kubiznak:2012wp, Mo:2014mba, Banerjee:2011au, Belhaj:2014tga}. The Ehrenfest equations describe the distinct second-order phase transitions of the vdW fluid as follows.
\begin{eqnarray}\label{ehrenfest1}
\left(\frac{\partial P}{\partial T}\right)_{S}=\frac{C_{P_2}-C_{P_1}}{T V(\alpha_{P_2}-\alpha_{P_1})}=\frac{\Delta C_{P}}{TV\Delta \alpha}
\end{eqnarray}
\begin{eqnarray}\label{ehrenfest2}
\left(\frac{\partial P}{\partial T}\right)_{V}=\frac{\alpha{_2}-\alpha{_1}}{\kappa_{T_2}-\kappa_{T_1}}=\frac{\Delta\alpha}{\Delta\kappa_T}
\end{eqnarray}
where $\alpha=\frac{1}{V}\left(\frac{\partial V}{\partial T}\right)_{P}$ and $\kappa_T=-\frac{1}{V}\left(\frac{\partial V}{\partial P}\right)_{T}$ are the isobaric volume expansion coefficient and the isothermal compressibility coefficient, respectively.\\
At critical points of $P$-$V$ criticality in the extended phase space thermodynamics, we analytically examine the Ehrenfest schemes described by Eqs. (\ref{ehrenfest1}) and (\ref{ehrenfest2}). The temperature, Eq.~(\ref{tempadsp}), as expressed in terms of entropy, can be obtained as follows.
\begin{eqnarray}\label{tementrop}
T&=&\frac{1}{4\pi}\Bigg[{(d-3)}\left(\frac{4S}{\Omega_{d-2}}\right)^{-\frac{1}{d-2}}+\frac{16\pi P}{(d(1-2\psi)-2)}\left(\frac{4S}{\Omega_{d-2}}\right)^{\frac{1}{d-2}}+{(3-d+\xi)}{N_s}\left(\frac{4S}{\Omega_{d-2}}\right)^{-\frac{1+\xi}{d-2}}\Bigg]\nonumber\\
\end{eqnarray}
Utilizing Eqs. (\ref{entropy}), (\ref{tempadsp}), and (\ref{tementrop}), we calculate the specific heat at constant pressure, the volume expansion coefficient, and the isothermal compressibility of the black holes as
\begin{eqnarray}
C_P&=& T\left(\frac{\partial S}{\partial T}\right)_{P}\nonumber\\
&=&\frac{(d-2)S\Bigg[1+\frac{16\pi P}{(d(1-2\psi)-2)(d-3)}\left(\frac{4S}{\Omega_{d-2}}\right)^{\frac{2}{d-2}}+\frac{{(3-d+\xi)}}{d-3}{N_s}\left(\frac{4S}{\Omega_{d-2}}\right)^{-\frac{\xi}{d-2}}\Bigg]}{\Bigg[-1+\frac{16\pi P}{(d(1-2\psi)-2)(d-3)}\left(\frac{4S}{\Omega_{d-2}}\right)^{\frac{2}{d-2}}-\frac{{(1+\xi)(3-d+\xi)}}{d-3}{N_s}\left(\frac{4S}{\Omega_{d-2}}\right)^{-\frac{\xi}{d-2}}\Bigg]}
\end{eqnarray}
\begin{figure*}
\includegraphics[scale=0.65]{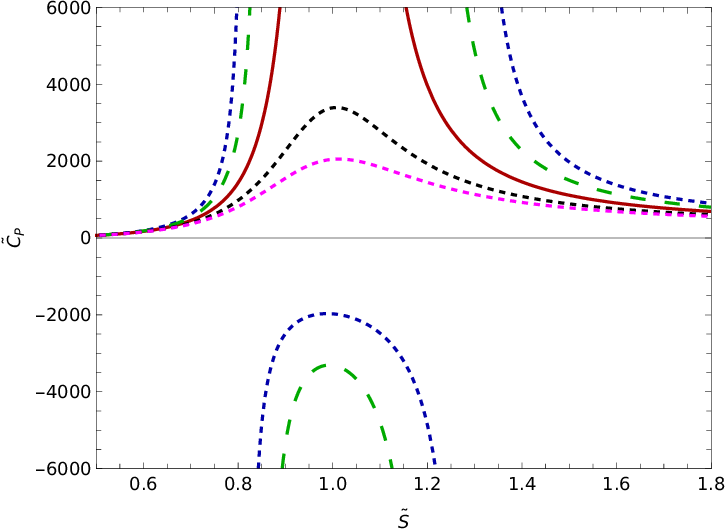}
\includegraphics[scale=0.65]{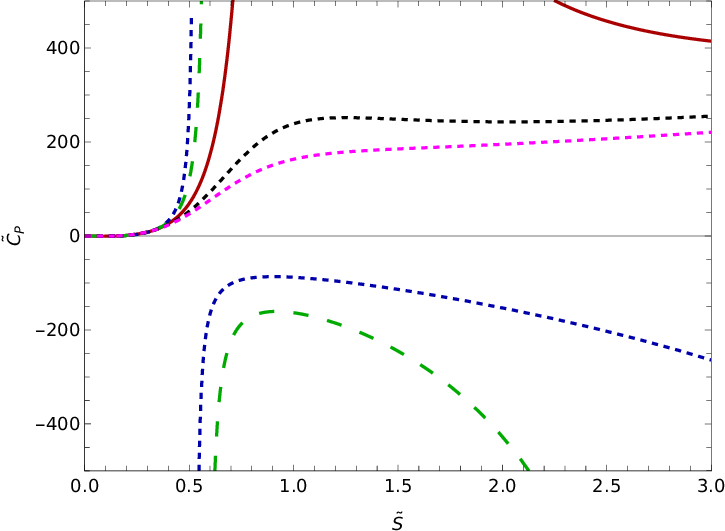}\\
\includegraphics[scale=0.65]{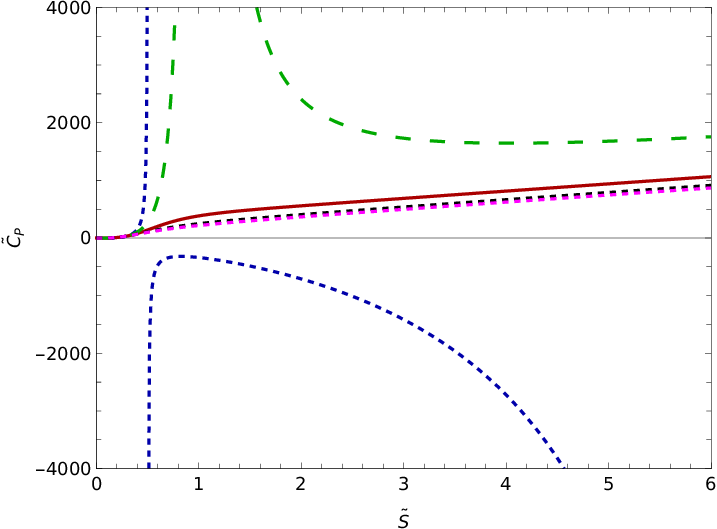}
\includegraphics[scale=0.65]{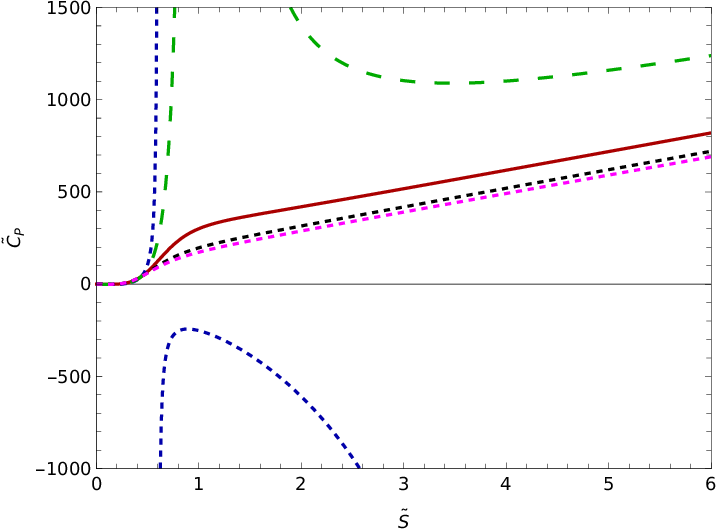}\\
\includegraphics[scale=0.65]{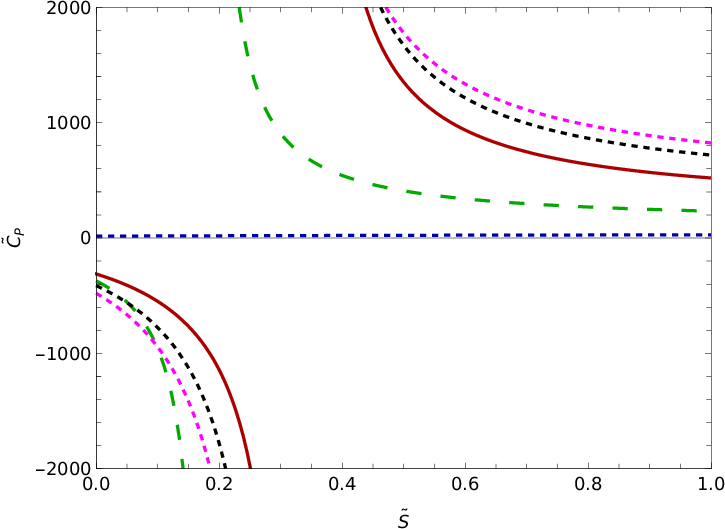}
\includegraphics[scale=0.65]{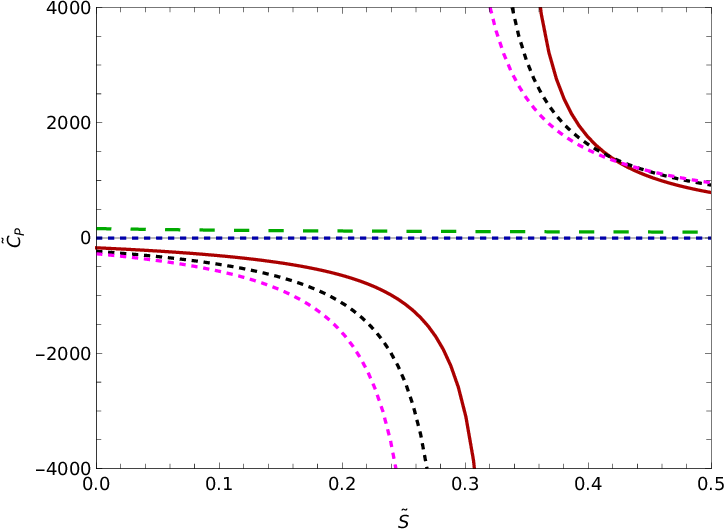}
 \caption{Plots showing the reduced heat capacity $\tilde{C}_P$ vs reduced entropy $\tilde{S}$ for the charged black hole in the AdS spacetime in Rastall gravity. \label{figCP}}
\end{figure*}
\begin{eqnarray}
\alpha &=&\frac{1}{V}\left(\frac{\partial V}{\partial T}\right)_{P}\nonumber\\
&=&\frac{4\pi\frac{d-1}{d-3}\left(\frac{4S}{\Omega_{d-2}}\right)^{\frac{2}{d-2}}}{\Bigg[-1+\frac{16\pi P}{(d(1-2\psi)-2)(d-3)}\left(\frac{4S}{\Omega_{d-2}}\right)^{\frac{2}{d-2}}-\frac{{(1+\xi)(3-d+\xi)}}{d-3}{N_s}\left(\frac{4S}{\Omega_{d-2}}\right)^{-\frac{\xi}{d-2}}\Bigg]}
\end{eqnarray}
\begin{figure*}
\includegraphics[scale=0.65]{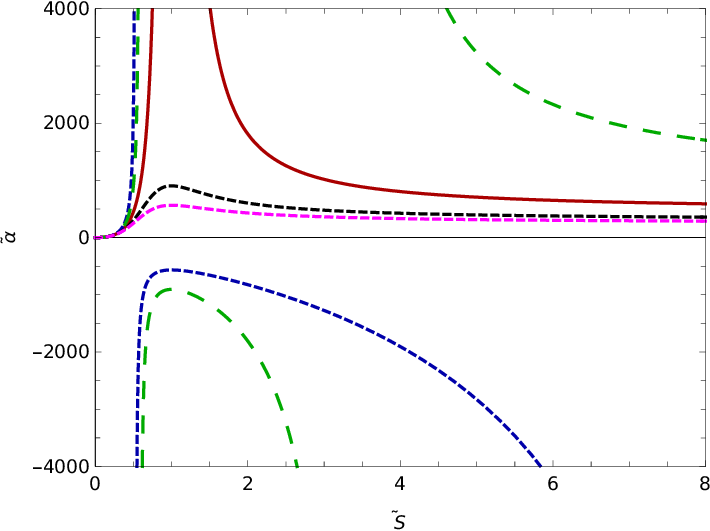}
\includegraphics[scale=0.65]{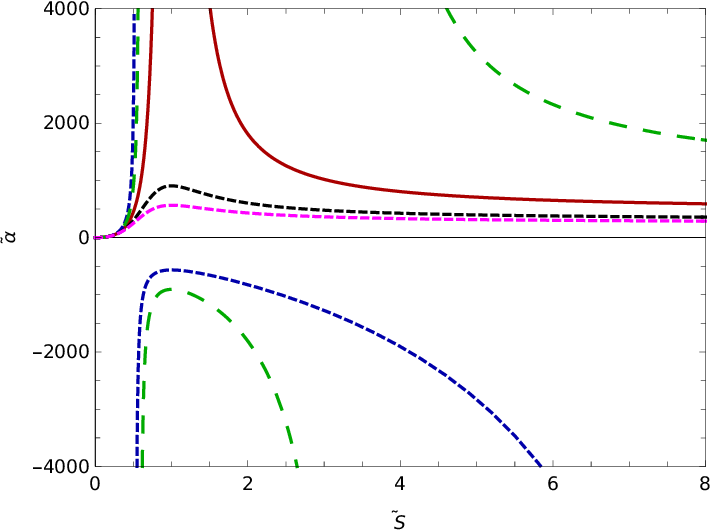}\\
\includegraphics[scale=0.65]{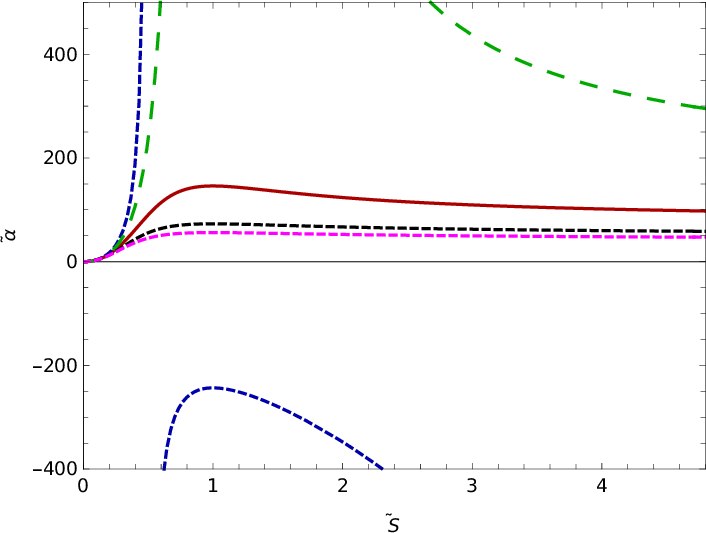}
\includegraphics[scale=0.65]{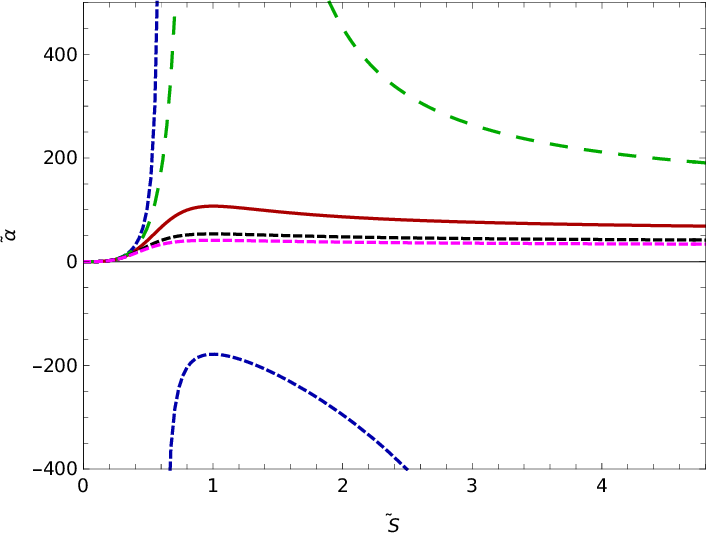}\\
\includegraphics[scale=0.65]{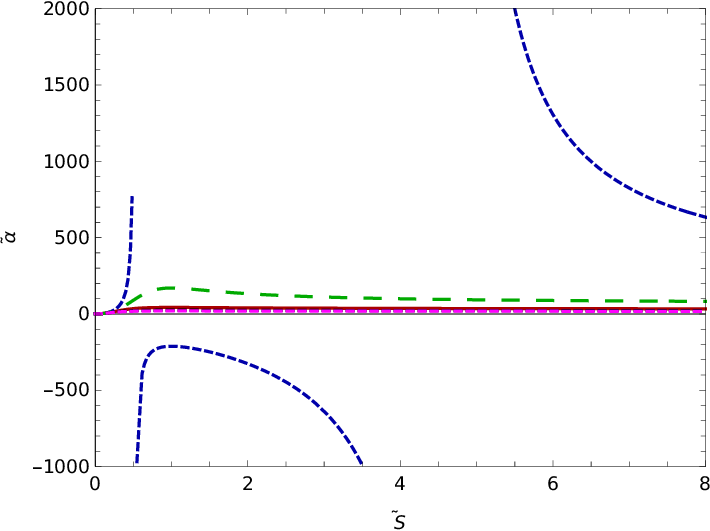}
\includegraphics[scale=0.65]{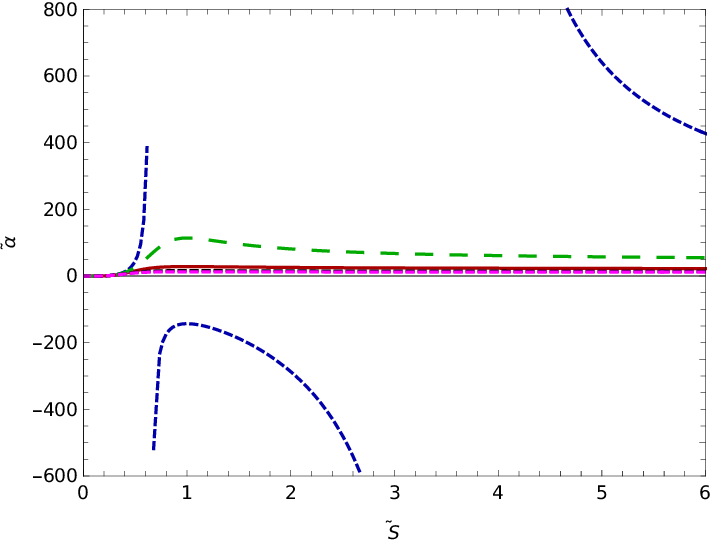}
 \caption{Plots showing the reduced volume expansion coefficient $\tilde{\alpha}$ vs reduced entropy $\tilde{S}$ for the charged black hole in the AdS spacetime in Rastall gravity. \label{figalpha}}
\end{figure*}
\begin{eqnarray}\label{kappaT}
\kappa_{T} &=&-\frac{1}{V}\left(\frac{\partial V}{\partial P}\right)_{T}\nonumber\\
&=&\frac{16\pi\frac{d-1}{(d(1-2\psi)-2)(d-3)}\left(\frac{4S}{\Omega_{d-2}}\right)^{\frac{2}{d-2}}}{\Bigg[-1+\frac{16\pi P}{(d(1-2\psi)-2)(d-3)}\left(\frac{4S}{\Omega_{d-2}}\right)^{\frac{2}{d-2}}-\frac{{(1+\xi)(3-d+\xi)}}{d-3}{N_s}\left(\frac{4S}{\Omega_{d-2}}\right)^{-\frac{\xi}{d-2}}\Bigg]}
\end{eqnarray}
\begin{figure*}
\includegraphics[scale=0.65]{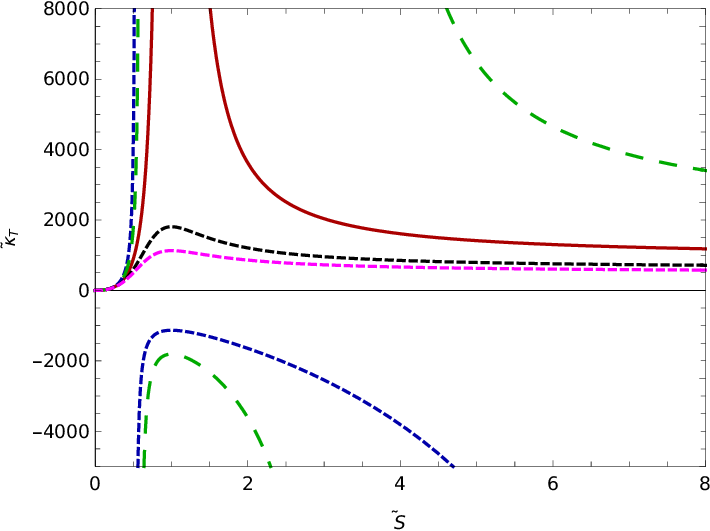}
\includegraphics[scale=0.65]{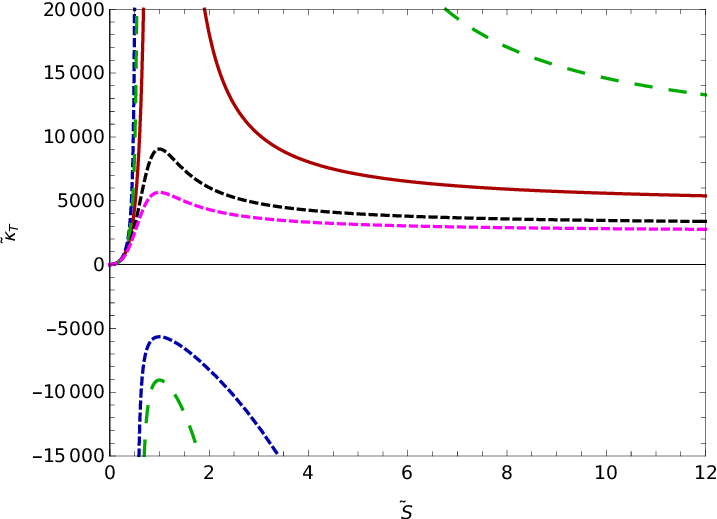}\\
\includegraphics[scale=0.65]{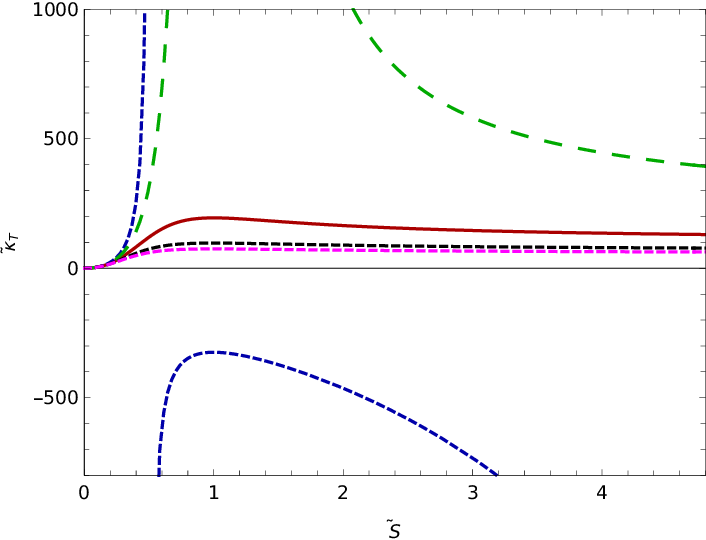}
\includegraphics[scale=0.65]{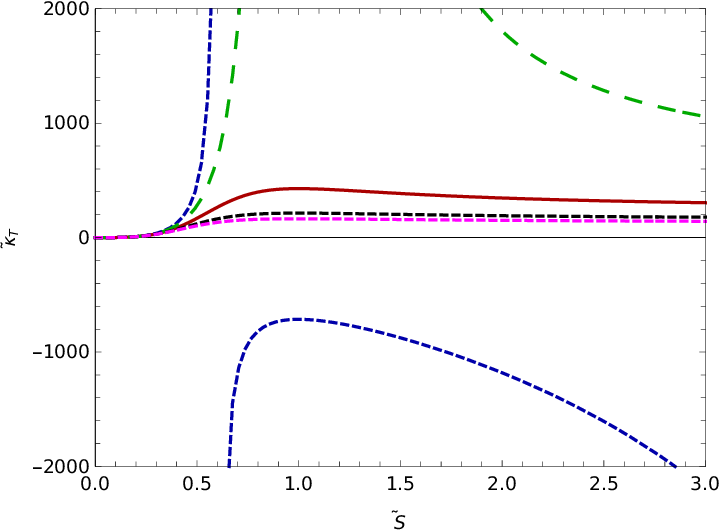}\\
\includegraphics[scale=0.65]{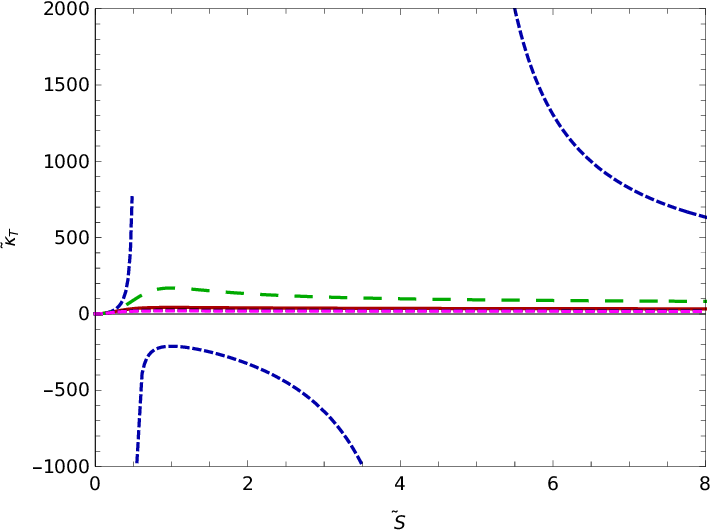}
\includegraphics[scale=0.65]{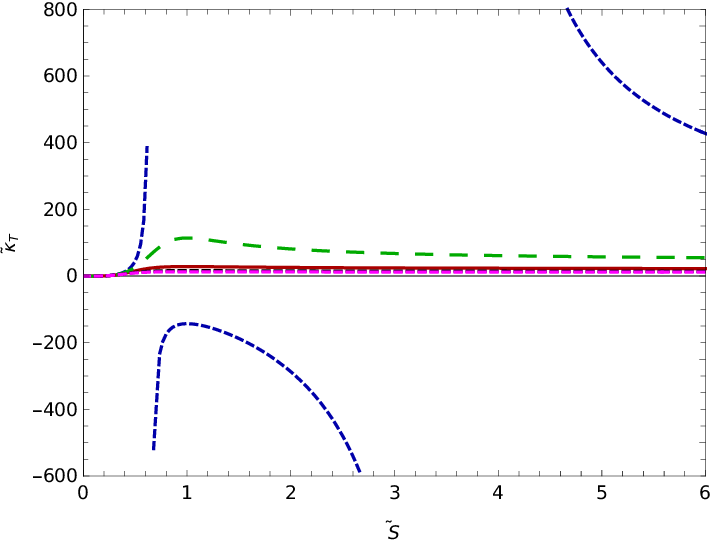}
 \caption{Plots showing the reduced isothermal compressibility $\tilde{\kappa}_T$ vs reduced entropy $\tilde{S}$ for the charged black hole in the AdS spacetime in Rastall gravity. \label{figkappa}}
\end{figure*}
In deriving the Eq.~(\ref{kappaT}), we use the following cyclic thermodynamic identity for a $(P,V,T)$ system 
\begin{eqnarray}\label{identity}
\left(\frac{\partial V}{\partial P}\right)_{T}\left(\frac{\partial P}{\partial T}\right)_{V}\left(\frac{\partial T}{\partial V}\right)_{P}=-1
\end{eqnarray}
It is to be mentioned that the denominators of $C_P$, $\alpha$, and $\kappa_T$ share a common factor, indicating that they diverge at the same point critical points. This leads to the condition defined as
\begin{eqnarray}\label{discc}
-1+\frac{16\pi P_c}{(d(1-2\psi)-2)(d-3)}\left(\frac{4S_c}{\Omega_{d-2}}\right)^{\frac{2}{d-2}}-\frac{{(1+\xi)(3-d+\xi)}}{d-3}{N_s}\left(\frac{4S_c}{\Omega_{d-2}}\right)^{-\frac{\xi}{d-2}}=0
\end{eqnarray}
Using Eqs. (\ref{entropy}), (\ref{r_h}) and (\ref{cvol}), we can obtain critical values of the quantities
\begin{eqnarray}\label{S_c}
S_c=\frac{\Omega_{d-2}}{4}\left[\frac{(3-d+\xi)(1+\xi)(2+\xi)N_s}{2(d-3)}\right]^{\frac{d-2}{\xi}}
\end{eqnarray}
\begin{eqnarray}\label{P_c}
P_c&=&\frac{(d(1-2\psi)-2)(d-3)\xi}{\pi (d-2)^2(2+\xi)}\left[\frac{(3-d+\xi)(1+\xi)(2+\xi)N_s}{2(d-3)}\right]^{\frac{-2}{\xi}}
\end{eqnarray} 
Therefore, Eqs.~(\ref{discc}) indicates that $C_P$, $\alpha$, and $\kappa_T$ exhibit discontinuities at the critical points $S_c$ and $P_c$. As depicted in Fig.~\ref{figCP}, we show the plots for the variation of reduced heat capacity $\tilde{C}_P$ concerning the reduced entropy $\tilde{S}$ for different values of reduced pressures. Similarly, in Figs.~\ref{figalpha} and \ref{figkappa}, we show the behaviors of reduced volume expansion coefficient $\tilde{\alpha}$ and reduced isothermal compressibility $\tilde{\kappa}_T$ as a function of reduced entropy for various values of the reduced pressures.\\
Next, we evaluate the validity of the Ehrenfest equations (\ref{ehrenfest1})-(\ref{ehrenfest2}) at the critical points. The definition of the volume expansion coefficient can be rearranged to yield
\begin{eqnarray}
V\alpha=\left(\frac{\partial V}{\partial T}\right)_{P}=\left(\frac{\partial V}{\partial S}\right)_{P}\left(\frac{\partial S}{\partial T}\right)_{P}=\left(\frac{\partial V}{\partial S}\right)_{P}\frac{C_P}{T},
\end{eqnarray}
Using Eqs.~(\ref{entropy}) and (\ref{vol1}), we can express the right-hand side of Eq.~(\ref{ehrenfest1}) as
\begin{eqnarray}\label{ehrencheck1}
\frac{\Delta C_{P}}{TV\Delta \alpha}=\left[\frac{d-2}{4}-\frac{\psi d}{2}\right]\left(\frac{S_c}{\Omega_{d-2}}\right)^{\frac{-1}{d-2}}
\end{eqnarray}
Applying Eq.~(\ref{tementrop}), we can derive the left-hand side of Eq. (\ref{ehrenfest1}) as
\begin{eqnarray}\label{ehrencheck2}
\Big[\left(\frac{\partial P}{\partial T}\right)_{S}\Big]_c=\left[\frac{d-2}{4}-\frac{\psi d}{2}\right]\left(\frac{S_c}{\Omega_{d-2}}\right)^{\frac{-1}{d-2}}
\end{eqnarray}
From Eqs. (\ref{ehrencheck1}) and (\ref{ehrencheck2}), we conclude that the first equation of the Ehrenfest relations is valid at the critical point.\\
Next, we calculate the second Ehrenfest equation as defined in Eq.~(\ref{ehrenfest2}) and check its validity. Using Eqs.~(\ref{entropy}), (\ref{volume1}) and (\ref{tementrop}), we find the left-hand side of Eq.~(\ref{ehrenfest2}) at the critical point can be expressed as 
\begin{eqnarray}
\left[\left(\frac{\partial P}{\partial T}\right)_V\right]_c=\left[\frac{d-2}{4}-\frac{\psi d}{2}\right]\left(\frac{S_c}{\Omega_{d-2}}\right)^{\frac{-1}{d-2}}
\end{eqnarray}
Using the defining relations of $\kappa_T$ and $\alpha$, we have 
\begin{eqnarray}
V\kappa_T=-\left(\frac{\partial V}{\partial P}\right)_T=\left(\frac{\partial T}{\partial P}\right)_V\left(\frac{\partial V}{\partial T}\right)_P=\left(\frac{\partial T}{\partial P}\right)_V V\alpha,
\end{eqnarray}
Thus, the right-hand side of Eq.~(\ref{ehrenfest2}) can be rewritten as
\begin{eqnarray}\label{ehrenfest22}
\frac{\Delta\alpha}{\Delta\kappa_T}=\left[\left(\frac{\partial P}{\partial T}\right)_V\right]_c=\left[\frac{d-2}{4}-\frac{\psi d}{2}\right]\left(\frac{S_c}{\Omega_{d-2}}\right)^{\frac{-1}{d-2}}
\end{eqnarray}
In deriving Eq.~(\ref{ehrenfest22}), once again we have employed the cyclic rule for the $P$-$V$-$T$ system defined in Eq.~(\ref{identity}). Consequently, the second Ehrefest equation is also valid at the critical points, as confirmed by Eq.~(\ref{ehrenfest22}). This indicates that both the Ehrenfest equations hold true at the critical point of the $P$-$V$ criticality of the $d$-dimensional Rastall AdS black hole surrounded by a quintessence field.\\
Using Eqs.~(\ref{ehrencheck1}) and (\ref{ehrenfest22}), we derive the Prigogine-Defay (PD) ratio as
\begin{eqnarray}\label{PDR}
\Pi=\frac{\Delta C_P\Delta \kappa_T}{TV\left(\Delta\alpha\right)^2}=1.
\end{eqnarray}
The PD ratio is identically equal to unity. Hence, it shows that the $d$-dimensional AdS black holes in a quintessence background within Rastall theory have a distinct second-order phase transition. The PD ratio was first introduced in \cite{prigogine:1954pd} and later on was extensively investigated in \cite{prabhat:1976pd}. This ratio can be used to measure the potential deviation from the system that does not show similar behavior as observed in vdW fluid \cite{Banerjee:2010qk, Banerjee:2010da}. In a vdW system, the PD ratio equals unity, while for a glassy phase transition, it ranges from 2 to 5 \cite{Banerjee:2010qk, Banerjee:2010da, Jacle:1986, Nieuwenhuizen:1997}. Eqs.~(\ref{PDR}) and the Ehrenfest relations enforcing the validity Eqs.~(\ref{ehrenfest1}) and (\ref{ehrenfest2}). It confirms a second-order phase transition at critical points. Therefore, for a $d$-dimensional AdS black holes in Rastall gravity, the phase transition is not an exception and follows the similar features of a vdW liquid-gas system.
\section{Off-shell Gibbs free energy}
\label{free_energy_LS}
\begin{figure}[h]
\includegraphics[scale=0.65]{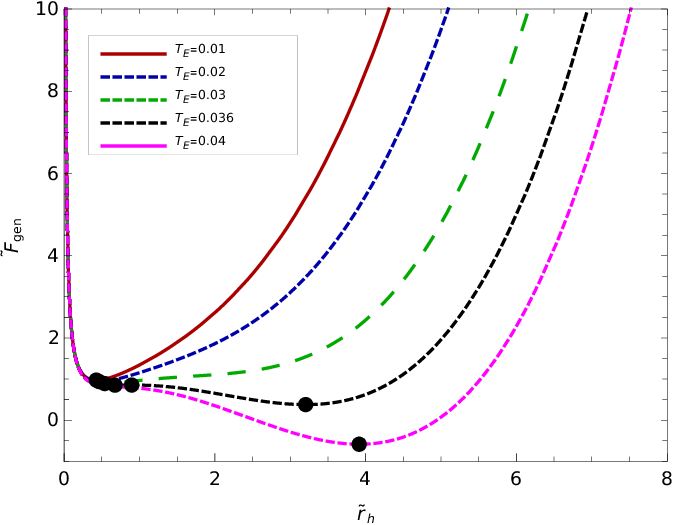}
\includegraphics[scale=0.65]{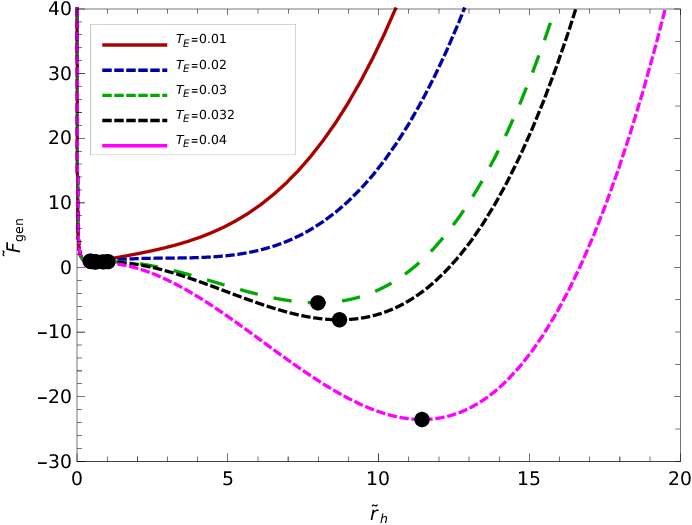}\\
\includegraphics[scale=0.65]{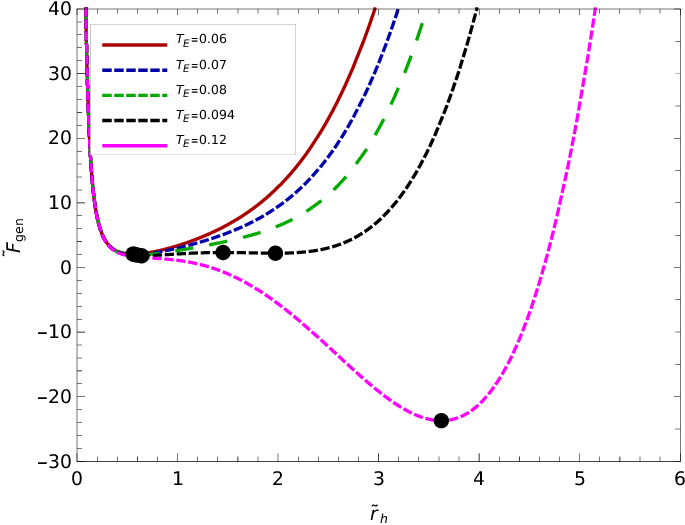}
\includegraphics[scale=0.65]{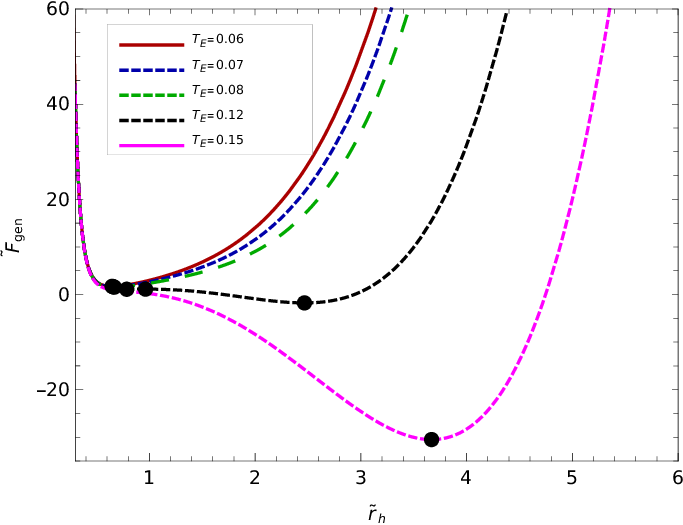}\\
\includegraphics[scale=0.65]{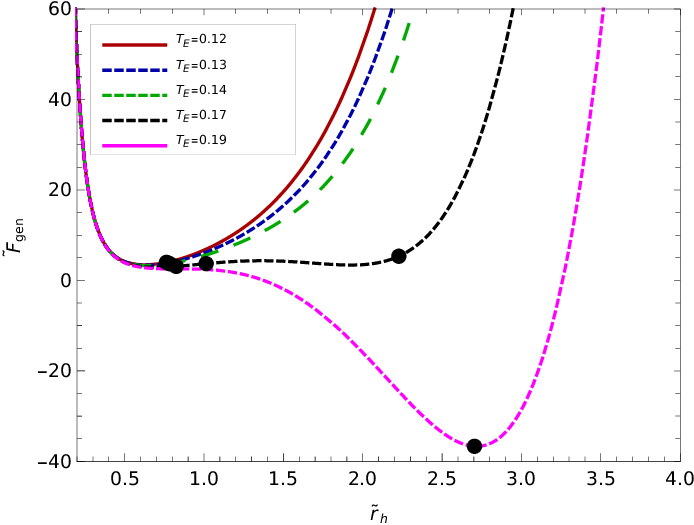}
\includegraphics[scale=0.65]{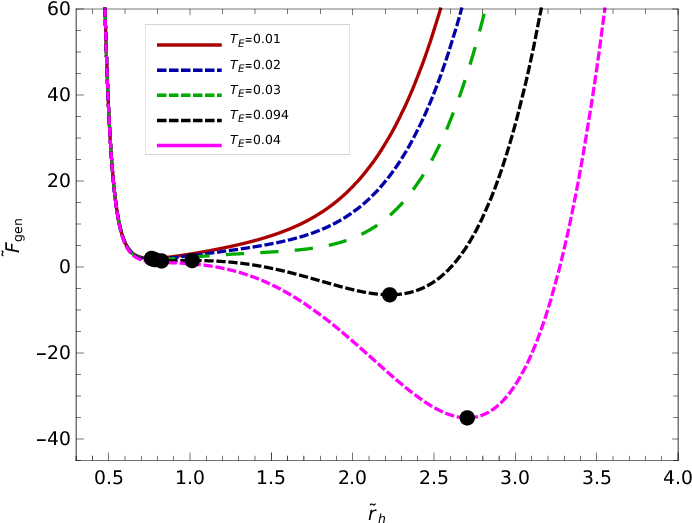}
 \caption{Plots showing the reduced off-shell Gibbs free energy $\tilde{F}_{\text{gen}}$ vs reduced event horizon $\tilde{r}_h$ for the charged black hole in the AdS spacetime in Rastall gravity. \label{figgenGibbs}}
\end{figure}

We qualitatively describe the off-shell Gibbs free energy across various spacetime dimensions for different values of the Rastall coupling parameter. The behavior of black hole phase transitions is better understood when we analyze the depth comprising the generalized free energy \cite{Yang:2021ljn}. The expression for the generalized free energy is written as
\begin{equation}
\label{genFEnergy}
F_\text{gen}=M-T_E S=\frac{(d-2)\Omega_{d-2}}{16\pi}\Bigg[1-\frac{N_s}{r_h^{\xi}}+\frac{16\pi\;P}{(d-1)(d(1-2\psi)-2)}r_h^2\Bigg]{r_h^{d-3}}-\frac{\Omega_{d-2}}{4}r_h^{d-2}T_E
\end{equation}
To qualitatively visualize the characteristics depths of the generalized free energy~(\ref{genFEnergy}), we depict in Fig.~\ref{figgenGibbs}, the reduced generalized free energy $\tilde{F}_{\text{gen}}$ versus the reduced horizon radius $\tilde{r}_h$ across different spacetime dimensions for various values of $\psi$. We illustrate these examples in connection to the Gibbs free energy $\tilde{F}_{\text{Gibbs}}$ of the black hole. To visualize the depths, we take as reference the reduced Gibbs free energy versus the reduced thermodynamic pressure for $\tilde{P}<1$ mimicking a distinctive swallow-tail behavior. We choose different ensemble temperatures that correspond to this swallow-tail diagram. The black points in the generalized off-shell Gibbs free energy, plotted against the horizon radius,  correspond to the equilibrium phases associated with the various ensemble temperatures. As shown in Figs.~\ref{figgenGibbs}, at different ensemble temperatures, we observe scenarios with single and double black hole phases, corresponding to the minima of the well. Such behaviors are evident in AdS black holes within Rastall theory in the presence of a quintessence field, applicable to any generic spacetime dimensions starting from four.
The extrema of the generalized free energy occur when the black hole is in thermodynamic equilibrium with its environment. The first law along with the generalized Gibbs free energy, we arrive at the following conditions.
\begin{equation}
    \begin{split}
        \left(\frac{\partial F_{\text{gen}}}{\partial r_h}\right)_{P,T_{\text{E}}} =\left[\left( \frac{\partial M}{\partial S}\right)_{P,T_{\text{E}}} - T_{\text{E}}\right]\left(\frac{\partial S}{\partial r_h}\right)_{P} = \left(T  - T_{\text{E}}\right)\left(\frac{\partial S}{\partial r_h}\right)_{P}.
    \end{split}
\end{equation}
It is of general perception that the thermal stability of the black hole is connected to the generalized free energy as follows, 
\begin{equation}
     \left.  \left( \frac{\partial^2 F_{\text{gen}}}{\partial r_h^2}\right)_{P,T_{\text{E}}}\right|_{T_{\text{E}}=T}=\left(\frac{\partial T}{\partial S}\right)_{P}\left(\frac{\partial S}{\partial r_h}\right)_{P}^2=\frac{T}{C_{P}}\left(\frac{\partial S}{\partial r_h}\right)_{P}^2,
\end{equation}
where $C_{P}$ is the heat capacity at constant pressure $P$. When the generalized free energy reaches its maximum, we have negative heat capacity, indicating the thermal instability of a black hole. On the other hand, the heat capacity becomes positive at the minimum of the generalized free energy, indicating local thermodynamic stability.
\section{Conclusions}\label{conclusion} 
In this paper, we discussed the $d$-dimensional AdS black holes in a quintessence field background in Rastall gravity. As limiting cases, when $\omega_s=-1\;\text{and}\;1/l^2\neq 0$, the metric (\ref{metricfunads}) reduced to the $d$-dimensional Schwarzschild-Tangherlini (A)dS black holes. We compared the new term in the metric, within the framework of Rastall theory of gravity, with the corresponding solution in a background quintessence field in general relativity, finding an effective equation of state parameter, $\omega_{eff}$. As limiting cases, we confronted our discussion for special solutions, such as those corresponding to a cosmological constant, a dust field, a quintessence background field, and a radiation field. Furthermore, we analyzed the horizon structures across various spacetime dimensions while considering different values of the Rastall coupling parameter. \\
As a precursor to incorporate the extended phase space thermodynamics, we considered cosmological constant as thermodynamic pressure, with its conjugate quantity identified as thermodynamic volume. We included the $VdP$ term in the first law of black hole mechanics. In addition to the $VdP$ term, the first law also incorporates the $\Theta_sdN_s$ term. We have studied the thermodynamics of $d$-dimensional AdS black hole spacetime in Rastall theory and defined a quantity $\Theta_s$ that is conjugate to $N_s$, ensuring consistency with the Smarr-Gibbs-Duhem relation. We focused on the investigation of the extended phase space thermodynamics of $d$-dimensional AdS black holes in a quintessence field background. The definitions of the thermodynamic pressure and volume term include the Rastall parameter $\psi$ and hence modified the thermodynamic quantities in AdS spacetime. We numerically presented the isobaric ($\tilde{P}-\tilde{v}$) and isothermal ($\tilde{T}-\tilde{v}$) curves in various spacetime dimensions for different values of the Rastall coupling parameters. To be more specific, we also have shown the extremal points on these curves by employing suitable extremality conditions.\\

We utilized the classical Ehrenfest schemes to examine the nature of phase transition at critical points of $P$-$V$ criticality. Our analysis focused on the classical Clausius-Clapeyron and Ehrenfest equations. We derived the heat capacity at constant pressure $C_P$, the isobaric volume expansion coefficient $\alpha$, and the isothermal compressibility coefficient $\kappa_T$. Using these thermodynamic quantities, we demonstrated that the Clausius-Clapeyron equations hold at critical points, thereby confirming the existence of first-order phase transitions. In addition, we analyzed the Ehrenfest relations and found they were also satisfied identically at the critical points, which confirmed the presence of second-order phase transitions. We also demonstrated the effect of the Rastall parameter $\psi$ in deriving these thermodynamic quantities in the extended phase space of black hole thermodynamics. Subsequently, we found that they diverge exactly at the same critical points. We calculated the PD ratio and found that it exactly equals unity. When $\psi\to 0$, our results are reduced to the corresponding expressions that were calculated in Einstein's general relativity. Consequently, the universal nature of the vdW liquid-gas system within the framework of Rastall gravity helped us to understand the inherent relationship of the thermodynamics of AdS black holes with the liquid-gas system. \\
We qualitatively described the off-shell Gibbs free energy across various spacetime dimensions for different values of the Rastall coupling parameter. The behavior of black hole phase transitions was better understood when we analyzed the depth comprising the generalized free energy. A classically feasible black hole solution corresponds to an extreme point in the generalized free energy versus the horizon diagram for different ensemble temperatures. In such diagrams, the maxima of generalized free energy led to unstable black hole solutions. When we found the minima in the generalized free energy, the black holes are likely stable.\\

Our results may be applied in the context of AdS/CFT correspondence. A rotating version of the black holes, along with their $P$-$V$ criticality investigation, will be physically motivated. Another possible future direction of our research could be a more rigorous exploration of the microscopic degrees of freedom. Among these include the molecular interactions of black hole micro-molecules during dynamical time evolution and kinetic turnover processes.

\end{document}